\documentclass[11pt]{article}

\usepackage[dvips]{graphicx}
\usepackage{epsfig,mathtools,amsmath,amssymb,verbatim,mathrsfs,array,layout,textcomp,latexsym, slashed,graphicx,bbm,physics,tensor,cite,caption}

\usepackage[normalem]{ulem}
\usepackage{appendix}
\usepackage{enumitem}
\usepackage{rotating}
\usepackage[hidelinks]{hyperref}
\usepackage[usenames,dvipsnames]{color} 

\usepackage{soul}
\usepackage{dsfont}

\usepackage{tikz}
\usepackage{tikz-feynman}
\tikzset{
	cross/.style={path picture={\draw[black]
			(path picture bounding box.south east) -- (path picture bounding box.north west)
			(path picture bounding box.south west) -- (path picture bounding box.north east);}}
}

\allowdisplaybreaks

\makeatletter

\newenvironment{bottompar}{\par\vspace*{\fill}}{\clearpage}
\newcommand*{\xdash}[1][3em]{\rule[0.5ex]{#1}{0.55pt}}
\def\mysection#1{{\bf #1.}}

\newcommand{\tg}{\tilde\gamma}
\newcommand\mydots{\hbox to 1.1em{$\,\cdot\hss\cdot\hss\cdot\,$}}

\begin{document}
	
	\begin{titlepage}
		\setcounter{page}{1} \baselineskip=15.5pt 
		\thispagestyle{empty}
		$\quad$
		{\raggedleft IFT UAM-CSIC 26-79\par}
		\vskip 60 pt
		
		\begin{center}
			{\fontsize{18}{18} \bf Background-independent one-loop renormalization \\ \vspace{0.15cm} 	of tensor and scalar primordial spectra}
		\end{center}
		
		\vskip 20pt
		\begin{center}
			\noindent
			{\fontsize{13}{30}\selectfont Guillermo Ballesteros$^{1,2}$, Jes\'us Gamb\'in Egea$^{2}$, and Flavio Riccardi$^{3,4}$}
		\end{center}
		
		\begin{center}
			\vskip 4pt
			\textit{ $^1${Departamento de F\'{\i}sica Te\'{o}rica, Universidad Aut\'{o}noma de Madrid (UAM), \\Campus de Cantoblanco, 28049 Madrid, Spain}
			}
			\vskip 5pt
			\textit{ $^2${Instituto de F\'{\i}sica Te\'{o}rica UAM-CSIC,  Campus de Cantoblanco, 28049 Madrid, Spain}
			}
			\vskip 5pt
			\textit{ $^{3}${Dipartimento di Fisica, Sapienza Università di Roma, Piazzale Aldo Moro 5, 00185, Roma, Italy}
			}
			\vskip 5pt
			\textit{ $^{4}${INFN sezione di Roma, Piazzale Aldo Moro 5, 00185, Roma, Italy}
			}
		\end{center}
		
		\vspace{0.4cm}
		\centerline{\bf Abstract}
		\vspace{0.3cm}
		\noindent 
		
		We present a background-independent renormalization framework for one-loop primordial spectra. We apply it to two observables: the scalar-induced tensor spectrum sourced by a minimally coupled spectator field, and the scalar spectrum generated by potential	self-interactions. In both cases, the UV part of the loops is isolated and, using the universal WKB behavior of the internal modes, the UV poles are extracted analytically and absorbed into local counterterms without having to specify either the background evolution or the full dynamics of the field running in the loop. This yields finite model-independent expressions for the renormalized spectra amenable to numerical analysis.
		
		\begin{bottompar}
			\noindent\xdash[15em]\\
			\small{
				guillermo.ballesteros@uam.es\\
				j.gambin@csic.es\\
				flavio.riccardi@uniroma1.it}
		\end{bottompar}
	\end{titlepage}
	
	\setcounter{page}{2}
	\newpage
	
	\tableofcontents
	
\section{Introduction} \label{sec: Intro}

Primordial correlation functions provide one of the main observational windows into the inflationary epoch. On cosmic microwave background (CMB) scales, the scalar fluctuations are observed to have a small amplitude (with an approximately flat spectrum) and to be very nearly Gaussian \cite{Planck:2018vyg,Planck:2019kim}. This justifies the standard description of these observables in terms of the free theory at tree level, with interactions treated perturbatively. Nevertheless, the study of the effects of interactions remains important and it is now very timely, as we argue below.
At tree level, interactions are probed by primordial non-Gaussianities \cite{Maldacena:2002vr,Bartolo:2004if,Chen:2010xka}. At loop level, they require understanding the renormalization \cite{Weinberg:2005vy,Senatore:2009cf} and physical interpretation of cosmological correlators \cite{Ballesteros:2025nhz}, and may become phenomenologically relevant in regimes where the fluctuations are enhanced.

The physical interpretation of loop corrections requires separating contributions that can be intrinsically assigned to loop diagrams from those that are tied to the freedom in the finite part of local counterterms. After the ultraviolet divergences have been absorbed, the finite remainder may still contain terms whose momentum dependence is identical to that generated by local counterterm insertions. Such terms 
are {\it scheme-dependent} and their numerical values can be modified by redefining the corresponding finite counterterm coefficients. Therefore they cannot be regarded as genuinely independent loop contributions. By contrast, a finite loop term is {\it distinguishable} only if its momentum dependence cannot be reproduced by local counterterms \cite{Ballesteros:2025nhz}. This distinction is especially important in inflation, where the time-dependent background and the coexistence of physical and comoving scales can obscure the separation between genuine loop effects and loops that are degenerate with counterterm contributions.

This issue has become particularly relevant in scenarios where slow-roll is transiently violated and the scalar spectrum is strongly enhanced, as in ultra slow-roll phases \cite{Tsamis:2003px,Kinney:2005vj,Martin:2012pe,Ballesteros:2017fsr}. The suggestion that enhanced scalar modes at small comoving scales could generate loop corrections large enough to spoil perturbation theory on CMB scales \cite{Kristiano:2022maq} triggered a broad discussion \cite{Kristiano:2022maq,Choudhury:2023vuj,Firouzjahi:2023bkt,Kristiano:2023scm,Riotto:2023gpm,Riotto:2023hoz,Firouzjahi:2023ahg,Franciolini:2023agm,Ballesteros:2024zdp}.
Part of the subsequent literature argued instead that, in ultra slow-roll models, the scalar loop contribution induced by scalar fluctuations vanishes identically on sufficiently large scales \cite{Fumagalli:2023zzl,Fumagalli:2024jzz,Tada:2023rgp,Inomata:2024lud,Ema:2026dop}. From the renormalization viewpoint, however, such a statement is necessarily scheme dependent (whenever the result is degenerate with local counterterms), and therefore cannot be assigned a well-defined physical meaning. Related questions, such as the role of tadpole renormalization in the conservation of superhorizon modes, have also been recently addressed \cite{Kristiano:2025ajj,Inomata:2026csq,Inomata:2025pqa}.

Dimensional regularization is one of the standard regularization methods in quantum field theory, in particular because it preserves the symmetry structure of the theory. In cosmological correlators, however, its implementation is technically involved. It is not enough to continue only the loop measure away from three spatial dimensions; the whole action must be continued consistently in order to preserve its symmetries, and so must, in particular, the mode functions and the interaction vertices \cite{Weinberg:2005vy,Senatore:2009cf}. The dynamics of the modes in extra dimensions complicates the (brute-force) analytic evaluation of the time integrals involved in the perturbative expansion of observables computed in the in-in formalism. The methods developed in \cite{Ballesteros:2024cef} and \cite{Ballesteros:2025nhz} make this problem tractable; see also \cite{Braglia:2025qrb,Kristiano:2025ajj,Braglia:2026fle,Fang:2026off} for applications in different inflationary contexts. The loop integral is split into an infrared (IR) part, evaluated directly in three spatial dimensions, and an ultraviolet (UV) part, treated in $3+\delta$ spatial dimensions, with $\delta$ the number of extra spatial dimensions. In the UV region, the universal WKB behavior of the modes running in the loop provides two complementary simplifications: it localizes the relevant regions of the time integrals contributing to the UV part, and it reduces the loop-momentum dependence to an asymptotic power series. The UV momentum integral can then be performed analytically, allowing the poles in $\delta$ to be extracted in a direct way. 

In this paper we use this framework to obtain general expressions for the renormalized one-loop primordial power spectra of tensor and scalar modes. The key observation that allows us to do this is the fact that the universal WKB expansion of the modes running in the loop makes it possible to isolate the UV contribution, including its $1/\delta$ divergence, directly from the asymptotic integrand. This can be done without specifying the background evolution (and therefore applies beyond strictly inflationary backgrounds) and without specifying the potential governing the fields in the loop. The divergent contribution can then be matched to counterterm insertions, while the renormalized observable is obtained by combining the finite infrared part, the finite ultraviolet remainder and the finite counterterm contribution. In this way, the procedure provides a model-independent prescription for renormalizing cosmological correlators.

We apply this strategy to two observables. First, we consider the tensor spectrum induced by the fluctuations of a minimally coupled spectator scalar field with arbitrary time-dependent mass, on a general FLRW background. This is especially interesting because it may dominate over the vacuum tensor signal without necessarily implying a breakdown of perturbation theory, since the latter is controlled solely by the free tensor action. We show that the UV divergences of this tensor one-loop spectrum are absorbed by counterterms descending from the covariant gravitational EFT. This is a non-trivial consistency check: for example, general covariance forbids a tensor mass counterterm, yet the covariant counterterm basis is sufficient to renormalize the induced tensor spectrum. The result is a finite expression that can be evaluated directly once the background evolution and the spectator dynamics are specified, making it particularly suitable for numerical applications since all analytic renormalization steps have already been performed.

Second, we study the scalar power spectrum generated by scalar self-interactions coming from a generic potential.\footnote{See also \cite{Braglia:2026fle} for a concrete one-loop renormalization analysis in an EFT of inflation setup allowing for arbitrary time-dependent coefficients.} These interactions dominate in regimes where the slow-roll parameter $\epsilon$ remains small while slow-roll is transiently violated through the evolution of higher-order slow-roll parameters, as in the ultra slow-roll examples mentioned above. In this regime, the interactions mediated by the algebraic metric variables are subleading, and the dominant scalar loop effects arise from the potential \cite{Maldacena:2002vr,Ballesteros:2024zdp}. Building on the role of tadpole renormalization in this context, clarified in \cite{Kristiano:2025ajj,Inomata:2026csq,Inomata:2025pqa}, 
we include the tadpole counterterm required to keep an arbitrary but fixed background evolution, and derive the corresponding renormalized scalar spectrum in closed form.
The final expression is finite and ready for direct application to models of interest. Moreover, it displays explicitly the dependence on the finite parts of the counterterms. Consequently, statements about the cancellation of the loop contribution on large (CMB) scales from transient slow-roll violations at smaller scales are scheme dependent.

These two applications clarify, in two complementary settings, how renormalization works for cosmological loop observables. The case of tensor modes shows how general covariance constrains the allowed counterterms in a time-dependent background. The scalar example shows how tadpole renormalization fixes the counterterm structure needed to renormalize the one-loop power spectrum in the potential-dominated regime. Taken together, they provide a concrete implementation of the general strategy and make the steps of the renormalization procedure explicit, so that the method can be adapted to other cosmological correlators.

The paper is organized as follows. In Section~\ref{sec: Dim Reg} we summarize the dimensional-regularization framework and the UV in-in simplification method. In Section~\ref{sec: SIGWs} we renormalize the tensor power spectrum induced by a spectator scalar and identify the covariant counterterms required for this purpose. In Section~\ref{sec: USR} we apply the same strategy to the scalar power spectrum generated by potential-induced self-interactions. We conclude in Section~\ref{sec: Discussion}. In Appendix~\ref{app: Cts action SIGWS} we derive the counterterm action involved in the renormalization of the tensor spectrum, while in Appendix~\ref{app: formulas} we collect useful intermediate expressions entering its derivation.

\section{Treatment of the cosmological loops} \label{sec: Dim Reg}

Consider the vacuum expectation value of a generic observable $\mathcal{O}(\tau)$, understood as a functional of the fields and their conjugate momenta, where $\tau$ denotes conformal time. In the in-in formalism \cite{Weinberg:2005vy,Schwinger:1960qe,Bakshi:1962dv,Bakshi:1963bn,Keldysh:1964ud}, this expectation value is written as
\begin{equation}
	\expval{\mathcal{O}(\tau)} = \bra{0} \overline{T}\exp(i \int_{-\infty_+}^\tau \dd \tau' H_I(\tau')) \, \mathcal{O}_{I}(\tau) \, T\exp(-i \int_{-\infty_-}^\tau \dd \tau' H_I(\tau')) \ket{0}\eval_{\rm n.b.} \,.
\end{equation}
Here n.b.\ stands for {\it no bubbles} \cite{Ballesteros:2024cef,Senatore:2016aui}, $T$ and $\overline{T}$ denote time and anti-time ordering, respectively, and the lower integration limits implement the standard $i\epsilon$ prescription through a small deformation of the time contour, $\tau_\pm = \tau(1\pm i \epsilon)$. Both $\mathcal{O}_{I}$ and the interaction Hamiltonian $H_I$ are evaluated in the interaction picture, where the fields evolve according to the free Hamiltonian. Expanding perturbatively in powers of $H_I$, we write
\begin{equation}
	\expval{\mathcal{O}(\tau)} \equiv \sum_{n = 0}^\infty \expval{\mathcal{O}(\tau)}^{(n)}
\end{equation}
with the first contributions given by
\begin{align} \label{eq: In-In short 1}
	\expval{\mathcal{O}(\tau)}^{(0)} =& \bra{0} \mathcal{O}_{I}(\tau) \ket{0} \,, \quad \expval{\mathcal{O}(\tau)}^{(1)} = 2 \Im{\int_{-\infty_-}^\tau \dd \tau' \, \bra{0} \mathcal{O}_{I}(\tau) H_I(\tau')\ket{0}} \,, \\ \nonumber \label{eq: In-In short 2}
	\expval{\mathcal{O}(\tau)}^{(2)} =& \int_{-\infty_+}^\tau \dd \tau' \int_{-\infty_-}^\tau \dd \tau'' \bra{0} H_I(\tau') \mathcal{O}_{I}(\tau) H_I(\tau'') \ket{0} \\
	&-2\Re{\int_{-\infty_-}^\tau \dd \tau' \int_{-\infty_-}^{\tau'} \dd \tau'' \bra{0}  \mathcal{O}_{I}(\tau) H_I(\tau') H_I(\tau'') \ket{0}} \,.
\end{align}
The $i\epsilon$ prescription provides the exponential damping that removes spurious contributions from the asymptotic past. Equivalently, it implements the projection required to compute expectation values in the interacting vacuum starting from the free vacuum in the remote past; see e.g.~\cite{Adshead:2009cb}.

A generic field in the interaction picture admits a decomposition in creation and annihilation operators determined by the free, quadratic Hamiltonian. Working in $d=3+\delta$ spatial dimensions, we write
\begin{equation}
	\phi(x) = \int \dfrac{\dd^{3+\delta} \vb{k}}{(2\pi)^{(3+\delta)/2}} e^{i \vb{k}\cdot \vb{x}} \phi_{\vb{k}}(\tau) \,, \quad  \phi_{\vb{k}}(\tau) = \phi_{k}(\tau) a_{\vb{k}}+ \phi_{k}^*(\tau) a^\dagger_{-\vb{k}} \quad {\rm with} \quad \comm{a_{\vb{k}}}{a^\dagger_{\vb{p}}} = \delta(\vb{k} - \vb{p})\,.
\end{equation}
The mode function $\phi_k(\tau)$ solves the free equation of motion in $3+\delta$ spatial dimensions, with asymptotic boundary conditions in $\tau\to -\infty$ selecting the positive-frequency solution \cite{Bunch:1978yq}. The corresponding free vacuum is defined by $a_{\vb{k}}\ket{0}=0$ and represents the asymptotic minimum-energy state, associated with the positive-frequency choice of the modes.

It is convenient to define a canonically normalized field $u(x)\equiv c(\tau)\phi(x)$, where $c(\tau)$ is a model-dependent function chosen so that the quadratic action for $u$ contains a canonical kinetic term. Its Fourier modes satisfy
\begin{equation} \label{eq: WKB eom}
	u''_k(\tau) + \left(k^2 + m_{\rm eff}^2(\tau) \right) u_k(\tau) = 0  \quad {\rm and} \quad \lim_{\tau \to - \infty} u_k(\tau) = \dfrac{e^{-i k \tau}}{\sqrt{2k}} \,,
\end{equation}
where primes denote derivatives with respect to conformal time $\tau$. We assume that, in the asymptotic past, the modes satisfy $k\gg \abs{m_{\rm eff}(\tau\to-\infty)}$, so that $u_k$ reduces to a massless Minkowski mode and the boundary condition above selects the positive-frequency solution.
Following the WKB approach \cite{Wentzel:1926aor,Kramers:1926njj,Brillouin:1926blg} (see also \cite{Weinberg:2010wq,Ballesteros:2025nhz}), the high-momentum behavior of the mode function with comoving momentum $k$ can be written as
\begin{equation} \label{eq: WKB}
	u_k(\tau) = \dfrac{e^{-i \int^\tau W_k(\tau') \dd \tau'}}{\sqrt{2 W_k(\tau)}}  \,,
\end{equation}
where the asymptotic positive-frequency condition has already been imposed. The function $W_k(\tau)$ admits an expansion at large $k$ given by
\begin{equation} \label{eq: WKB frec}
	W_k = k + \dfrac{m_{\rm eff}^2}{2k} - \dfrac{m_{\rm eff}^4 + \left( m_{\rm eff}^2\right) ''}{8k^3} + \dfrac{2m_{\rm eff}^6 + 5\left( \left( m_{\rm eff}^2\right) '\right) ^2 + 6 m_{\rm eff}^2 \left( m_{\rm eff}^2\right) '' +\left( m_{\rm eff}^2\right) ''''}{32k^5} + \order{k^{-7}} \,.
\end{equation}
This expansion is reliable provided that the effective mass varies adiabatically on the scale set by $k$, namely $\abs{\dd^n m_{\rm eff}^2(\tau)/\dd\tau^n}\ll k^{n+2}$ for all $n\in \mathbb{N}$.

Let us anticipate that the existence of the asymptotic positive-frequency regime,
$u_k(\tau \to -\infty)\propto e^{-ik\tau}$, is the only model-dependent requirement that will be needed in the renormalization of the spectra studied below. This condition constrains the effective mass and, ultimately, the background evolution and the dynamics of the fields. Accordingly, the renormalized spectra derived in Sections~\ref{sec: SIGWs}--\ref{sec: USR} rely only on the assumption that the internal modes running in the loop admit the asymptotic behavior captured by the WKB expansion above.

\subsection{Dimensional regularization}

Loop corrections to cosmological observables computed with the perturbative in-in expansion are generally ultraviolet divergent and must therefore be regularized. A regulator is simply an auxiliary modification of the UV behavior of the theory that renders intermediate expressions finite; physical observables cannot depend on its choice \cite{Feynman:1948fi,Tomonaga:1948zz,Schwinger:1949zz,Dyson:1949bp}. This dependence is removed by renormalization, through local counterterms that absorb the UV-regulator dependence of the loop corrections. We will discuss this procedure explicitly in later sections.

The regulator we will use is dimensional regularization \cite{Bollini:1972bi,tHooft:1972tcz,Cicuta:1972jf,Ashmore:1972uj} (see e.g.~\cite{Weinberg:2005vy,Senatore:2009cf} for inflationary applications), which is particularly convenient because it preserves the symmetry structure of the theory. Its consistent implementation in inflationary correlators was discussed in \cite{Senatore:2009cf}, and the practical method that we use was developed in \cite{Ballesteros:2024cef}. A generic one-loop contribution to an observable can be written in the form
\begin{equation}
	I(\delta) = \int _0^\infty \dd p \, p^\delta f(\delta,p)\,.
\end{equation}
The factor $p^\delta$ isolates the explicit effect of the Fourier-space volume element in $3+\delta$ spatial dimensions and serves as the regulator that ensures UV convergence. All the remaining information is contained in $f(\delta,p)$: the interaction vertices, the external and internal mode functions, the angular integrations, and the time integrals characteristic of the in-in formalism. In particular, $f(\delta,p)$ also contains the implicit dependence on $\delta$ induced by continuing the theory away from three spatial dimensions.

The key observation is that only the UV part of the integral needs to be treated in $3+\delta$ spatial dimensions. Assuming that no IR divergences are present, we split the integral at an arbitrary comoving scale $L$,
\begin{equation} \label{eq: decomp}
	I(\delta) =  \int _0^L \dd p \, f(0,p) + \int _L^\infty \dd p \, p^\delta f(\delta,p)+\mathcal{O}(\delta)\,.
\end{equation}
The first term is finite by construction and can therefore be evaluated directly in three spatial dimensions. The second term may contain UV divergences and must be kept at $\delta\neq0$. The scale $L$ is arbitrary, but it is useful to choose it much larger than the comoving scales of the problem, so that the UV contribution can be extracted from the large-$p$ expansion of the integrand. The final result cannot depend on $L$; this independence is recovered through the cancellation between the explicit $L$-dependence of the two terms in Eq.~(\ref{eq: decomp}).

Let us now assume that, at large loop momentum, $f(\delta,p)$ admits an asymptotic Laurent expansion with a highest power $N$,
\begin{equation}
	f(\delta,p) = \sum_{n = -\infty}^N p^n \, c_n(\delta) \,,
\end{equation}
where the coefficients $c_n(\delta)$ may depend on the external momenta and on time and are analytic around $\delta=0$. This is the UV behavior expected for inflationary loop diagrams with Bunch--Davies (positive frequency) initial conditions, as follows from the high-momentum expansion of the modes in Eq.~(\ref{eq: WKB}), see \cite{Ballesteros:2025nhz}. More exotic UV structures can be treated with the same logic \cite{Ballesteros:2024cef}.

The UV integral is first evaluated in the domain of complex $\delta$ where it is convergent, and the resulting expression is then analytically continued to a neighborhood of $\delta=0$. Expanding the analytically continued result around $\delta=0$, one finds
\begin{equation} \label{eq: Dim Reg First Paper}
		I(\delta) = \lim_{L\to \infty} \left( \int_0^L \dd p \, f(0,p) - c_{-1}(0) \log L - \sum_{n = 0}^N \dfrac{L^{n+1}}{n+1} c_n(0) \right) - \dfrac{c_{-1}(0)}{\delta} - \dfrac{\dd c_{-1}(\delta)}{\dd \delta} \eval_{\delta = 0} + \order{\delta} \,.
\end{equation}
The expression inside the $\lim_{L \to \infty}$ is the three-dimensional loop integral with its logarithmic and power-law UV divergences explicitly subtracted. Dimensional regularization then adds two characteristic terms. The first one, $-c_{-1}(0)/\delta$, is the pole associated with the logarithmic divergence. The second one, $-\dd c_{-1}/\dd\delta\vert_{\delta=0}$, is the finite contribution generated by the continuation of the integrand away from three dimensions. This term is essential: it is the only place where the non-trivial $\delta$-dependence of $f(\delta,p)$ contributes to the finite UV part under the assumptions stated above.

Equation~(\ref{eq: Dim Reg First Paper}) also makes clear why, in practice, one only needs the integrand up to first order in $\delta$. Since the UV expansion considered here produces at most a simple pole at $\delta=0$, terms of order $\delta^2$ in $f(\delta,p)$ cannot contribute to the finite part after the limit $\delta\to0$ is taken, and the computation of the UV contribution reduces to identifying the coefficient of $p^{-1}$ in the large-$p$ expansion of both $f(0,p)$ and $\partial_\delta f(\delta,p)\vert_{\delta=0}$.

\subsection{Ultraviolet loop treatment}

The dimensional-regularization prescription described above concerns the loop-momentum integral. In cosmological observables, however, loop integrals appear inside the in-in expansion and are therefore accompanied by time integrals. In dimensional regularization, the continuation away from three spatial dimensions modifies not only the momentum measure, but also the mode functions and interaction vertices entering the integrand, which can complicate the associated time integrals. As a result, the UV part of the loop generally involves time integrals over genuinely $3+\delta$-dimensional quantities.

These time integrals can nevertheless be simplified in a systematic way, as explained in \cite{Ballesteros:2025nhz}. The point is that the scale $L$ introduced in Eq.~(\ref{eq: decomp}) to separate the IR and UV regions is arbitrary and can be taken much larger than any other comoving scale. In the UV region, $p>L$, the modes running inside the loop can then be treated in their high-momentum regime. Their time dependence is controlled by the universal WKB expansion discussed above, see Eq.~(\ref{eq: WKB}), and this will allow us to simplify the time integrals appearing in the loop contributions.

For one-loop diagrams with a single interaction vertex this already leads to a direct simplification. For instance, in a scalar loop generated by a non-derivative interaction, the internal contraction gives a momentum integral over $\abs{\phi_p(\tau')}^2$. In terms of the canonically normalized WKB mode, this contribution is proportional to $1/W_p(\tau')$, whose large-$p$ expansion follows immediately from Eq.~(\ref{eq: WKB frec}). The UV part of the loop integral can then be computed directly with the method of the previous section, by extracting the relevant powers of $p$ in the large-momentum expansion.

The simplification is more powerful for loops involving two interaction vertices. To illustrate it, consider a representative scalar loop generated by non-derivative interactions. Starting from Eq.~(\ref{eq: In-In short 2}), the corresponding contribution can be written in Fourier space, following \cite{Ballesteros:2025nhz}, as
\begin{align} \label{eq:genericdiagram}
	\begin{tikzpicture}[baseline={-2}]
		\draw (-0.15,0.3+0.15) -- (0.5,0);
		\fill (0,0.15) circle (1pt);
		\fill (0,0) circle (1pt);
		\fill (0,-0.15) circle (1pt);
		\draw (-0.15,-0.3-0.15) -- (0.5,0);
		\draw (0.5+0.25,0) circle (0.25);
		\draw (1,0) -- (1.5+0.15,0.3+0.15);
		\fill (1.5,0.15) circle (1pt);
		\fill (1.5,0) circle (1pt);
		\fill (1.5,-0.15) circle (1pt);
		\draw (1,0) -- (1.5+0.15,-0.3-0.15);
	\end{tikzpicture} = \int & \dfrac{\dd^{3+\delta} \vb{p}}{(2\pi)^{3+\delta}} \bigg[ \mathcal{I}_1\left(\tau,{\vb{p}};\{\vb{k}_i\};\delta\right)+\mathcal{I}_2\left(\tau,{\vb{p}};\{\vb{k}_i\};\delta\right)\bigg]\,,
\end{align}
with
\begin{align} \label{int1}
	\mathcal{I}_1\left(\tau,{\vb{p}};\{\vb{k}_i\};\delta\right)  & =  \int_{-\infty_+}^ \tau \dd \tau' \int_{-\infty_-}^ \tau \dd \tau'' \, G_1(\tau,\tau',\tau'';\{\vb{k}_i\};\delta) \, \phi_p(\tau') \phi_p^*(\tau'') \phi_q(\tau') \phi_q^*(\tau'')\,, \\  \label{int2}
	\mathcal{I}_2\left(\tau,{\vb{p}};\{\vb{k}_i\};\delta\right)  & = - 2\Re{  \int_{-\infty_-}^ \tau \dd \tau' \int_{-\infty_-}^ {\tau'} \dd \tau'' \, G_2(\tau,\tau',\tau'';\{\vb{k}_i\};\delta) \, \phi_p(\tau') \phi_p^*(\tau'') \phi_q(\tau') \phi_q^*(\tau'')} \,.
\end{align}
Here $\vb{k}_i$ denotes the external momenta and the two loop momenta are $\vb{p}$ and $\vb{q}$, with $\vb{q} = \vb{k}_{\rm tot} - \vb{p}$, where $\vb{k}_{\rm tot}$ is the total external momentum entering one of the vertices. The functions $G_1$ and $G_2$ collect the dependence on the external legs, the vertices and possible time-dependent couplings. 
We have chosen a non-derivative interaction and a single scalar field running in the loop only to simplify the notation; the same logic applies to derivative interactions and multi-field observables, see \cite{Ballesteros:2025nhz} for the details.

Given the WKB form of the mode functions in Eq.~(\ref{eq: WKB}), the UV expansion of the loop integrand contains a universal rapidly oscillating phase, $e^{-i (p+q) (\tau'-\tau'')}$, multiplying an asymptotic series in inverse powers of the loop momenta. It is therefore convenient to factor out this phase and define
\begin{equation} \label{eq: Def G two-vertex}
	G_i(\tau,\tau',\tau'';\{\vb{k}_i\};\delta) \, \phi_p(\tau') \phi_p^*(\tau'') \phi_q(\tau') \phi_q^*(\tau'') \equiv e^{-i(p+q)(\tau'-\tau'')} \mathcal{G}_i(\tau,\tau',\tau'';\{\vb{k}_i\},{\vb{p}};\delta)\,,\quad  i =1,2 \,.
\end{equation}
The $i\epsilon$ prescription damps the rapidly oscillating phase through the deformation of the integration contour into the complex plane. As a result, in the large-$p$ expansion, the only boundary regions that contribute are $\tau'=\tau''=\tau$ for $\mathcal{I}_1$ and $\tau''=\tau'$ for the nested integral $\mathcal{I}_2$. Keeping the phase unexpanded, one can then Taylor expand the slowly varying factor $\mathcal{G}_i$ around these regions, obtaining
\begin{align} \label{eq: Dim Reg Ints Tempo 2 Vert I}
	&\mathcal{I}_1\left(\tau,{\vb{p}};\{\vb{k}_i\};\delta\right)  = \sum_{{n,m} = 0}^{{\infty}} \partial^n_{\tau''} \partial^m_{\tau'}\, \mathcal{G}_1(\tau,\tau',\tau'';\{\vb{k}_i\},{\vb{p}};\delta) \eval_{\tau' = \tau'' = \tau} \dfrac{{i^{n-m}}}{(p+q)^{2+n+m}} \,,\\ \label{eq: Dim Reg Ints Tempo 2 Vert II}
	&\mathcal{I}_2\left(\tau,{\vb{p}};\{\vb{k}_i\};\delta\right)  = -2 {\rm Re} \left\{ \int_{-\infty_-}^ \tau \dd \tau' \sum_{n = 0}^\infty \partial^n_{\tau''}\,  \mathcal{G}_2(\tau,\tau',\tau'';\{\vb{k}_i\},{\vb{p}};\delta) \eval_{\tau'' = \tau'} \dfrac{{i^{n-1}}}{(p+q)^{1+n}}\right\}\,.
\end{align}
These expressions should be understood as UV asymptotic expansions. Their usefulness comes from inserting them only in the UV region $p>L$, where corrections suppressed by powers of $1/L$ can be discarded, and only a finite number of terms in the sums contribute. After the phase has been extracted, the remaining factors $\mathcal{G}_i$ contain the loop-momentum dependence only through an asymptotic power series in $p$ (and $q$). The momentum integral is therefore elementary, while the non-trivial time dependence is confined to the coefficients of this expansion. We refer to \cite{Ballesteros:2025nhz} for the detailed derivation. This simplification relies only on the universal UV phase of the free modes and on the $i\epsilon$ prescription, and therefore applies equally to derivative interactions.

Although the same strategy can be generalized to diagrams with more insertions of $H_I$, in this work the analysis of one-loop diagrams with one and two interaction vertices will be sufficient.

\subsection{Renormalization strategy}

The procedure explained in the previous subsections provides a direct route to renormalize one-loop observables before specifying the details of the model under consideration. After splitting the loop integral into IR and UV contributions, the UV part is treated with the high-momentum WKB expansion of the internal modes. Once the universal UV phase has been isolated and the time integrals simplified, the loop-momentum dependence reduces to an asymptotic power series whose coefficients collect the non-universal information left after this factorization. The momentum integral can then be performed analytically, allowing the UV poles in $1/\delta$ to be extracted in a model-independent way.

For one-loop diagrams, the divergent UV contribution contains at most one remaining time integral. Schematically, the pole in $1/\delta$ takes the form of a time integral over the external mode functions, multiplied by an effective time-dependent coupling fixed by the WKB coefficients of the fields running in the loop. This is the key simplification: the divergent part of the loop is identified before evaluating the remaining time integral, and therefore before specifying the background evolution and the detailed dynamics of the fields involved.

The same structure is generated by local counterterms, which contribute through tree-level diagrams with the same external fields and the same type of time integral. The UV loop contribution and the counterterm contribution can therefore be matched directly at the level of the integrand. The renormalization strategy is thus reduced to three steps: isolate the UV part of the loop and apply the UV in-in expansion; perform the resulting asymptotic loop-momentum integral to extract the pole in $1/\delta$; and fix the divergent part of the counterterm coefficients by cancelling this pole. The renormalized observable is then obtained by combining the finite IR contribution, the finite UV remainder and the finite counterterm contribution.

\section{Scalar-induced gravitational waves} \label{sec: SIGWs}

In this section we compute the one-loop tensor spectrum sourced by the fluctuations of a minimally coupled spectator scalar field. The spectator assumption means that this field does not determine the homogeneous geometry, which we take to be a fixed FLRW background. We keep the background evolution arbitrary, not necessarily inflationary, and encode the scalar dynamics relevant for the loop through a general time-dependent mass. We will show that the ultraviolet divergences of the induced tensor spectrum are absorbed by local counterterms compatible with the covariant gravitational EFT. This leads to a renormalized expression that can be evaluated directly for a generic  background evolution and any dynamics of the spectator field. 
Finally, we provide a concrete example for a de Sitter displaced-mass case.

\subsection{Scalar--tensor action}

We consider a spectator scalar field $\overline{\chi}(x)$ minimally coupled to gravity,
\begin{equation} 
	S = \int \dd^{4+\delta} x \, \mu^\delta \,\sqrt{-g} \, \left( \dfrac{M_P^2}{2} R - \dfrac{1}{2}g^{\mu\nu} \partial_\mu \overline{\chi} \partial_\nu \overline{\chi} - V(\overline{\chi})\right) \,.
\end{equation}
Although the energy density carried by $\overline{\chi}$ is assumed to be negligible for the background evolution, its coupling to the metric sources loop corrections to tensor correlators. Our goal is to compute the corresponding one-loop contribution to the tensor two-point function. To this end, we need the quadratic actions governing the free scalar and tensor modes, together with the scalar--tensor interaction vertices.

Since the loop calculation will be regularized in dimensional regularization, we work in $3+\delta$ spatial dimensions. The factor $\mu^\delta$, where $\mu$ has dimensions of energy, compensates the extra-dimensional volume element, so that the canonical mass dimensions of the fields remain independent of $\delta$. Moreover, on a FLRW background the measure $\dd^{4+\delta}x\sqrt{-g}$ is invariant under a constant rescaling of the scale factor, $a\to \lambda a$, accompanied by the corresponding rescaling of comoving coordinates, $\{ \tau,\vb{x}\} \to \lambda^{-1} \, \{ \tau,\vb{x}\}$, see \cite{Senatore:2009cf}. Consequently, invariance of the full action under this symmetry requires $\mu$ to be a physical energy scale, namely $\mu\to\mu$, rather than a comoving scale transforming as $E\to \lambda E$. This  strongly constrains the structure of the loop-level results \cite{Senatore:2009cf,Ballesteros:2025nhz}.

We use the ADM decomposition of the metric \cite{Arnowitt:1962hi,Maldacena:2002vr},
\begin{equation} \label{eq: ADM ds2}
	\dd s^2 = -N^2 \dd t^2 + \gamma_{ij} (N^i \dd t + \dd \vb{x}^i) (N^j \dd t + \dd \vb{x}^j) \,, \quad {\rm with} \quad \gamma_{ij} = a^2 \tilde \gamma _{ij}= a^2 \left( e^\Gamma\right) _{ij} \,,
\end{equation}
where the relation between cosmic time, $t$, and conformal time is $\dd t = a \dd \tau$.
We then decompose the metric perturbations into scalar, vector and tensor components as
\begin{equation} \label{eq: ADM SVT}
	\Gamma_{ij} = 2\zeta \delta_{ij} +\partial_{ij} E + \partial_{(i} E_{j)} + h_{ij} \,,
\end{equation}
where $\partial_i E_i=0$, while $h_{ij}$ is transverse and traceless,
$\partial_i h_{ij} = h_{ii} = 0$. The lapse $N$ and the shift $N_i$ are non-dynamical variables fixed by constraint equations. In what follows we restrict the analysis to the scalar-induced contribution in the spectator regime. In this regime, the mixing between the spectator and the constraint variables is suppressed by the small contribution of the spectator field to the background stress-energy, parametrically controlled by its background velocity \cite{Maldacena:2002vr,Cheung:2007st,Weinberg:2008hq}. We therefore neglect the scalar--tensor interactions mediated by the lapse and shift, retaining only the direct minimal-coupling vertices.

Let us write the spectator field as a homogeneous background plus a fluctuation, $\overline{\chi}(x) \equiv \chi_0(t) + \chi(x)$. The expansion of the scalar potential gives
\begin{equation}
	V(\overline{\chi}) = V(\chi_0) + V'(\chi_0) \, \chi + \dfrac{1}{2} V''(\chi_0) \, \chi^2 + \order{\chi^3} \,.
\end{equation}
The quadratic term defines the time-dependent mass of the spectator fluctuation, $m_\chi^2(\tau) \equiv V''(\chi_0(\tau))$. 
This is the only information about the spectator potential that enters the one-loop tensor spectrum in the setup considered here. We keep $m_\chi^2(\tau)$ arbitrary, without specifying its time dependence.

With these considerations, and neglecting scalar--tensor interactions mediated by the lapse and shift as discussed above, we work in a gauge in which the scalar sector is carried entirely by the spectator fluctuation $\chi$ and retain only the transverse-traceless tensor modes $h_{ij}$ in the metric. The action relevant for the one-loop tensor spectrum is then
\begin{align} \label{eq: SIGWs Action}
	\nonumber
	S = \int \dd \tau \, \dd^{3+\delta} \vb{x} \, \mu^\delta a^{2+\delta} \Bigg\lbrace &\dfrac{M_P^2}{8} \left(h'^2_{ij} - (\partial_k h_{ij})^2  \right) +\dfrac{1}{2}\left(\chi'^2-(\partial_i \chi)^2 - a^2(\tau) \, m_\chi^2(\tau) \chi^2 \right) \\
	&- \dfrac{1}{2} \left( - h_{ij} + \dfrac{1}{2} h_{ik}h_{kj} + \order{h_{ij}^3} \right) \partial_i \chi \partial_j \chi  \Bigg\rbrace \,.
\end{align}
For the present computation it is sufficient to keep the quadratic tensor and scalar sectors, together with the scalar--tensor interactions up to fourth order in fluctuations. These are precisely the vertices entering the one-loop tensor two-point function.

\subsubsection{Counterterm action}

As usual, the action in Eq.~(\ref{eq: SIGWs Action}) must be supplemented by local counterterms whose insertions in the tensor two-point function generate diagrams at the same perturbative order as the scalar one-loop contribution. Their role is to absorb the UV divergences of the loop diagrams and render the tensor two-point function finite. We follow the covariant EFT construction of \cite{Ballesteros:2024cef}, keeping only the ingredients needed here and extending the analysis to an arbitrary FLRW background and a spectator scalar with general time-dependent mass.

The interactions in Eq.~(\ref{eq: SIGWs Action}) generate two one-loop topologies: a diagram with two cubic insertions and a diagram with a single quartic insertion,
\begin{equation}
	\expval{h_{ij}(x)h_{ij}(y)}^{\rm one-loop} = 
	\begin{tikzpicture}[baseline={-2}]
		\draw[decorate, decoration={snake, amplitude=-2pt, segment length=10pt}] (3,0) -- (5,0);
		\draw (4,0.58) circle (0.5);
		\fill[black] (3,0) circle (1.5pt);
		\fill[black] (5,0) circle (1.5pt);
		\node[black] at (3,-0.3) {$x$};
		\node[black] at (5,-0.3) {$y$};
		
		\node[black] at (5.5,0) {$+$};
		
		\draw[decorate, decoration=snake] (6,0) -- (7,0);
		\draw[decorate, decoration=snake] (8,0) -- (9,0);
		\draw (7.5,0) circle (0.5);
		\fill[black] (6,0) circle (1.5pt);
		\fill[black] (9,0) circle (1.5pt);
		
		\node[black] at (6.5,0.3) {$h_{ij}$};
		\node[black] at (7.5,0.7) {$\chi$};
	\end{tikzpicture} \propto \frac{1}{M_P^4} \,.
\end{equation}
Both contributions have the same scaling in $M_P$. Indeed, the quadratic tensor action is proportional to $M_P^2$, so each interaction-picture tensor mode function carries a factor $M_P^{-1}$, whereas the scalar mode functions and the minimal-coupling vertices in Eq.~(\ref{eq: SIGWs Action}) introduce no additional powers of $M_P$. The counterterm contribution must therefore have the same schematic scaling,
\begin{equation}
	\expval{h_{ij}(x)h_{ij}(y)}^{\rm counterterms} = 
	\begin{tikzpicture}[baseline={-2}]
		\draw[decorate, decoration=snake] (0,0) -- (2,0);
		\draw[fill=white,cross] (1,0) circle (0.2);
		\fill[black] (0,0) circle (1.5pt);
		\fill[black] (2,0) circle (1.5pt);
	\end{tikzpicture}
	\propto \frac{1}{M_P^4}\,,
\end{equation}
and corresponds to a quadratic insertion in the tensor sector. Since the required $M_P^{-4}$ suppression is already supplied by the tensor normalization, the counterterm coefficients relevant for this renormalization problem must be dimensionless.

We determine the allowed counterterms from the EFT of gravity viewpoint \cite{Donoghue:1994dn,Ruhdorfer:2019qmk}, by classifying the covariant operators that can contribute to the quadratic action for tensor fluctuations. Since we are interested only in the tensor two-point function, it is enough to keep the tensor dependence of the ADM variables. Tensor modes can enter through $\gamma_{ij}$, $N$ and $N_i$, but scalar, vector and tensor perturbations decouple at linear order. Hence, tensor-dependent corrections to the lapse and shift start at $\order{h_{ij}^2}$ and do not contribute to the quadratic tensor action \cite{Maldacena:2002vr,Wang:2013zva}. We may therefore set $N=1$ and $N_i=0$ when extracting the tensor counterterms, while keeping the tensor fluctuations in $\gamma_{ij}$.

The required $M_P$ scaling selects four-derivative curvature invariants with dimensionless coefficients. In the present spectator setup, there is one additional dimension-two background quantity, associated with the curvature of the spectator potential. A convenient covariant basis (see e.g.~\cite{Weinberg:2008hq,Weinberg:2010wq}) is therefore
\begin{equation} \label{eq:Cts action general}
	S = \int \sqrt{-g} \,\dd^{4+\delta}x\, \mu^\delta \, \big\{  A_1 R^2 + A_2 R_{\mu\nu}R^{\mu\nu} + A_3 R_{\mu\nu\rho\sigma}R^{\mu\nu\rho\sigma}  + A_4 \Box R + \tilde{V}''(\overline{\chi}) R \big\}\,.\footnote{Counterterms involving derivatives of the spectator background, such as $\partial_\mu \overline{\chi}$, together with curvature invariants are higher-order operators in the EFT and are therefore not required for the one-loop tensor-spectrum renormalization considered here.}
\end{equation}
The coefficients $A_i$ are dimensionless. The last operator extends the purely gravitational basis used in \cite{Ballesteros:2024cef}.\footnote{In \cite{Ballesteros:2024cef}, no spectator potential was included and the scalar field was massless; see Eq.~(\ref{eq: SIGW coef cts}) below, where the divergent part of $\tilde m_\chi^2$ vanishes for $m_\chi^2=0$.} Once evaluated on the spectator background, $\tilde{V}''(\chi_0)$ behaves as a time-dependent mass-squared and gives a dimension-four operator without introducing additional powers of $M_P$. Higher-curvature operators suppressed by powers of $M_P$ are not required to renormalize the one-loop tensor spectrum.

Expanding Eq.~(\ref{eq:Cts action general}) to quadratic order in $h_{ij}$, and using the free tensor equations of motion to remove redundant operators, gives
\begin{align} \label{eq: Cts Action SIGWs}
	\nonumber
	S = &\int \dd \tau \dd^{3+\delta} \vb{x} \, \mu^\delta \, a^{2+\delta} \, \Bigg\lbrace \left[ C_1 \left( \dfrac{1}{2}\dot H + H^2\right) + \dfrac{\tilde m_\chi^2(\tau)}{4}\right]  \left( h'^2_{ij} -  (\partial_k h_{ij})^2\right) + C_2 \dot H \left( h'^2_{ij} + \dfrac{1}{3} (\partial_k h_{ij})^2\right) \\ \nonumber
	&+C_3\left[\dfrac{1}{a^2}\left( (\partial^2h_{ij})^2 -  (\partial_k h'_{ij})^2\right) + \dot H h'^2_{ij} \right] +C_4\left(\dfrac{2}{3}\dot H + H^2 \right) (\partial_kh_{ij})^2  \\
	&+  \delta \, \dot H \,  \left[   - \dfrac{C_1}{8} \left( h'^2_{ij} - (\partial_k h_{ij})^2\right) -\dfrac{C_3}{2} h'^2_{ij}  +\dfrac{C_4 - C_2}{9}(\partial_kh_{ij})^2 \right] \Bigg\rbrace + \order{\delta} \,.
\end{align}
Here $H$ denotes the Hubble function and its derivative with respect to cosmic time is denoted as $\dot H$. The coefficients $C_i$ are dimensionless linear combinations of the $A_i$ in Eq.~(\ref{eq:Cts action general}), chosen so as to minimize the set of independent quadratic operators, and overdots denote derivatives with respect to cosmic time $t$. The effect of the non-minimal coupling $\tilde V''(\bar\chi)$  has been denoted as
\begin{equation}
	\tilde m_\chi^2(\tau) \equiv \tilde V''(\chi_0(\tau))\,.
\end{equation}
Spatial boundary terms have been discarded, and the temporal boundary terms that do not contribute to the tensor power spectrum have been removed \cite{Arroja:2011yj, Burrage:2011hd, Braglia:2024zsl}; the detailed expansion is given in Appendix~\ref{app: Cts action SIGWS}.

The terms proportional to $\delta$ in Eq.~(\ref{eq: Cts Action SIGWs}) must be retained. Even though counterterm diagrams contain no loop integrals, their coefficients absorb the UV poles of the loop correction and therefore contain simple poles in $\delta$. Consequently, the $\order{\delta}$ terms generated by expanding the covariant operators can leave finite remnants when multiplied by these poles. The explicit extra-dimensional corrections displayed in Eq.~(\ref{eq: Cts Action SIGWs}) are proportional to $\dot H$, and therefore vanish in exact de Sitter. 

\subsubsection{Fields in the interaction picture}

We now specify the interaction-picture fields entering the in-in computation of the tensor spectrum. Working in dimensional regularization, with $d=3+\delta$ spatial dimensions, we decompose the scalar and tensor fluctuations as
\begin{equation}
	\chi(\tau,\vb{x}) = \int \frac{\dd^{3+\delta}\vb{k}}{(2\pi)^{3/2}} e^{i\vb{k}\cdot\vb{x}} \chi_{\vb{k}}(\tau) \quad {\rm and} \quad h_{ij} (\tau,\vb{x}) = \int \frac{\dd^{3+\delta}\vb{k}}{(2\pi)^{3/2}} e^{i\vb{k}\cdot\vb{x}} \sum_{\gamma} e^\gamma_{ij}(\vb{k}) h_{\vb{k}}^\gamma(\tau) \,.
\end{equation}
We keep the Fourier prefactor at its three-dimensional value, $(2\pi)^{-3/2}$. The extra-dimensional effect associated with the corresponding overall factor in loop integrals is degenerate with local counterterms --it is scheme-dependent-- and can therefore be removed without loss of generality \cite{Ballesteros:2024cef}.

The index $\gamma$ labels the tensor polarizations in $d$ spatial dimensions. The only polarization identities needed for the induced tensor spectrum are \cite{Cardoso:2002pa,Ballesteros:2024cef}
\begin{equation} \label{eq: SIGWs polarization prop}
	\sum_\gamma e^\gamma_{il}(\vb{k}) e^\gamma_{lj}(\vb{k}) \vb{p}_i \vb{p}_j = \frac{(d-2)(d+1)}{2(d-1)} p^2\sin^2\theta_{\vb{p},\vb{k}} \quad {\rm and} \quad \sum_\gamma \left( e^\gamma_{ij}(\vb{k}) \vb{p}_i \vb{p}_j\right)^2 = \frac{d-2}{d-1}\left( p^2\sin^2\theta_{\vb{p},\vb{k}}\right) ^2,
\end{equation}
where $\theta_{\vb{p},\vb{k}}$ is the angle between $\vb{p}$ and $\vb{k}$.

In the interaction picture, the fields are expanded in creation and annihilation operators as
\begin{equation}
	\chi_{\vb{k}}(\tau) = \chi_k(\tau) a_{\vb{k}} + \chi_k^*(\tau) a^\dagger_{-\vb{k}} \quad {\rm and} \quad h_{\vb{k}}^\gamma(\tau) = h_k(\tau) a_{\vb{k}}^\gamma + h_k^*(\tau) a_{-\vb{k}}^{\gamma,\dagger} \,.
\end{equation}
The tensor mode function $h_k(\tau)$ carries no polarization label because the free tensor dynamics is polarization independent, see Eq.~(\ref{eq: SIGWs Action}).

The free equations are most conveniently written in terms of the canonically normalized variables $\chi_{c,k}$ and $h_{c,k}$, defined from Eq.~(\ref{eq: SIGWs Action}) as
\begin{align} \label{eq: SIGWs eom scalar}
	& \chi''_{c,k} +  \left( k^2 + m_{{\rm eff},\chi}^{2}(\tau)\right) \chi_{c,k}  = 0  \quad {\rm with} \quad \chi_{c,k} \equiv \mu^{\tfrac{\delta}{2}} a^{\tfrac{2+\delta}{2}} \chi_k \,, \\ \label{eq: SIGWs eom tensor}
	& h''_{c,k} +  \left( k^2 + m_{{\rm eff},h}^{2}(\tau)\right) h_{c,k}  = 0  \quad {\rm with} \quad h_{c,k} \equiv \dfrac{M_P}{2} \mu^{\tfrac{\delta}{2}} a^{\tfrac{2+\delta}{2}} h_k \,.
\end{align}
The corresponding effective masses are
\begin{align}
	&m_{{\rm eff},\chi}^{2} \equiv a^2\left[ m_\chi^2 - \dfrac{2+\delta}{2} \dot H - \dfrac{4+3\delta}{2} H^2  \right] +\order{\delta^2}\,, \\
	&m_{{\rm eff},h}^{2} \equiv a^2\left[ - \dfrac{2+\delta}{2} \dot H - \dfrac{4+3\delta}{2} H^2  \right] +\order{\delta^2} \,.
\end{align}
As discussed in Section~\ref{sec: Dim Reg}, dimensional regularization only requires the extra-dimensional corrections up to $\order{\delta}$, so terms of order $\order{\delta^2}$ and higher can be consistently neglected.

The creation and annihilation operators in the mode expansion define the free vacuum $\ket{0}$. Requiring this state to coincide with the asymptotic minimum-energy state fixes the positive-frequency solutions of the canonically normalized modes, 
\begin{equation} \label{eq: SIGWs asymp exp}
	\lim_{\tau \to - \infty}  \chi_{c,k} \sim \dfrac{e^{-i k \tau}}{\sqrt{2k}} \quad {\rm and} \quad \lim_{\tau \to - \infty}  h_{c,k} \sim \dfrac{e^{-i k \tau}}{\sqrt{2k}} \,.
\end{equation}
This prescription requires a regime in the far past where $k^2\gg m_{\rm eff}^2$ and the plane-wave behavior solves the free equations and the canonical commutation relations are recovered. The validity of this asymptotic regime is the only assumption needed in the renormalization procedure below.

The canonically normalized equations for the modes are of the form Eq.~(\ref{eq: WKB eom}). Their large-$k$ WKB solutions can then be written as
\begin{equation} \label{eq: SIGWs WKB}
	\chi_{c,k}(\tau) =  \dfrac{ e^{-i \int^\tau_{\tau_*} W_k^\chi(\tau') \dd \tau'}}{\sqrt{2W_k^{\chi}(\tau)}} \quad {\rm and } \quad h_{c,k}(\tau) =  \dfrac{ e^{-i \int^\tau_{\tau_*} W_k^h(\tau') \dd \tau'}}{\sqrt{2W_k^{h}(\tau)}} \,,
\end{equation}
where the frequencies $W_k^{\chi,h}(\tau)$ admit the large-momentum expansion given in Eq.~(\ref{eq: WKB frec}), with the corresponding effective masses. The reference time $\tau_*$ fixes only an overall phase, but it is useful to keep it when manipulating the UV time integrals.

\subsection{One-loop tensor power spectrum}

We now compute the one-loop contribution to the tensor two-point function sourced by the spectator scalar. The interaction Hamiltonian in the interaction picture, including both the scalar--tensor vertices in Eq.~(\ref{eq: SIGWs Action}) and the counterterms in Eq.~(\ref{eq: Cts Action SIGWs}), reduces at the perturbative order considered here to $H_I=-L_I$,\footnote{For the scalar--tensor vertices this is immediate, since they contain no time derivatives of the fields. Counterterms do contain tensor time derivatives, but the corrections to the interaction Hamiltonian obtained by identifying $H_I$ with $-L_I$ in the interaction picture are suppressed by additional powers of $M_P^{-1}$; see e.g.~\cite{Pimentel:2012tw,Arroja:2011yj,Ballesteros:2025nhz}.} where $L_I$ is the interaction Lagrangian in the interaction picture:
\begin{align} \label{eq: SIGWs HI}
	\nonumber
	H_I(\tau) =& \int \dd^{3+\delta}\vb{x} \, \mu^\delta a^{2+\delta} \Bigg\{-\frac{1}{2} h_{ij} \partial_i \chi \partial_j \chi + \frac{1}{4} h_{il} h_{lj} \partial_i \chi \partial_j \chi \\ \nonumber
	& -\left[ C_1 \left( \dfrac{1}{2}\dot H + H^2\right) + \dfrac{\tilde m_\chi^2(\tau)}{4}\right]  \left( h'^2_{ij} -  (\partial_k h_{ij})^2\right) - C_2 \dot H \left( h'^2_{ij} + \dfrac{1}{3} (\partial_k h_{ij})^2\right) \\ \nonumber
	&-C_3\left[\dfrac{1}{a^2}\left( (\partial^2h_{ij})^2 -  (\partial_k h'_{ij})^2\right) + \dot H h'^2_{ij} \right] -C_4\left(\dfrac{2}{3}\dot H + H^2 \right) (\partial_kh_{ij})^2  \\
	&-  \delta \, \dot H \,  \left[   - \dfrac{C_1}{8} \left( h'^2_{ij} - (\partial_k h_{ij})^2\right) -\dfrac{C_3}{2} h'^2_{ij}  +\dfrac{C_4 - C_2}{9}(\partial_kh_{ij})^2 \right] \Bigg\} \,,
\end{align}
where all fields are understood to be in the interaction picture.

At any perturbative order, momentum conservation and isotropy imply that the {\it connected} tensor two-point function can be written as
\begin{equation}\label{eq:Def Power Spectrum}
	\expval{h_{ij}(\tau,\vb{x}) h_{ij}(\tau,\vb{y})}_c = \int \frac{\dd^{3+\delta}\vb{k}}{(2\pi)^3} e^{i\vb{k}\cdot (\vb{x} - \vb{y})} \frac{2\pi^2}{k^{3+\delta}} \mathcal{P}_h(\tau,k) \times 2\,.
\end{equation}
The factor $\times 2$ is included so that, in three spatial dimensions, $\mathcal{P}_h(\tau,k)$ coincides with the usual dimensionless power spectrum per tensor polarization. We again use a Fourier convention with an overall $(2\pi)^{-3}$ factor, since the extra-dimensional effects associated with this choice are scheme dependent; by contrast, the factor $k^\delta$ is intrinsic to dimensional regularization and must be kept to preserve the correct dimensionality \cite{Ballesteros:2024cef}. Disconnected one-point contributions do not affect $\mathcal{P}_h$, and moreover isotropy implies $\expval{h_{ij}}=0$.

At one loop, there are two types of scalar-induced contributions. The first one comes from a single insertion of the quartic vertex and, using Eq.~(\ref{eq: In-In short 1}), gives
\begin{equation} \label{eq: SIGWs P q}
	\mathcal{P}_h^{\rm q}(\tau,k) =\frac{k^{3+\delta}}{4\pi^2} \int  \frac{\dd^{3+\delta} \vb{p}}{(2\pi)^3} \left(1+\dfrac{3\delta}{4} \right)  p^2 \sin^2\theta  \Im{h_{k}^2 \int_{-\infty_-}^\tau \dd \tau' \, h_{k}^{*2} \abs{\chi_{c,p}}^2}\,.
\end{equation}
The second one comes from two insertions of the cubic vertex and, using Eq.~(\ref{eq: In-In short 2}), can be written as
\begin{align}
	&\mathcal{P}_h^{\rm c}(\tau,k) \equiv \mathcal{P}_{h}^{\rm c,1}(\tau,k) + \mathcal{P}_{h}^{\rm c,2}(\tau,k) \,, \quad {\rm with} \label{eq: SIGWs P c} \\
	&\mathcal{P}_h^{\rm c,1}(\tau,k) = \frac{k^{3+\delta}}{4\pi^2} \int\frac{\dd^{3+\delta}\vb{p}}{(2\pi)^3} \left(\frac{1}{2}+\dfrac{\delta}{4} \right)  p^4 \sin^4\theta \abs{h_k}^2  \int_{-\infty_+}^\tau \mkern-10mu \dd \tau' \, h_k \chi_{c,p} \chi_{c,q} \int_{-\infty_-}^\tau \mkern-10mu \dd \tau'' \, h^*_k  \chi_{c,p}^*  \chi_{c,q}^* \,, \\
	&\mathcal{P}_h^{\rm c,2}(\tau,k) = \frac{k^{3+\delta}}{4\pi^2} \int\frac{\dd^{3+\delta}\vb{p}}{(2\pi)^3} \left(\frac{1}{2} +\dfrac{\delta}{4} \right)  p^4 \sin^4\theta  \Re \bigg\{ -2h_k^2 \int_{-\infty_-}^\tau \mkern-10mu \dd \tau' \, h^*_k \chi_{c,p} \chi_{c,q}  \int_{-\infty_-}^{\tau'} \mkern-10mu \dd \tau'' \, h_k^*   \chi_{c,p}^* \chi_{c,q}^* \bigg\}\,.
\end{align} 
Here we used the polarization identities in Eq.~(\ref{eq: SIGWs polarization prop}).
The internal spectator modes have been canonically normalized, see Eq.~(\ref{eq: SIGWs eom scalar}). Hence, the factor $\mu^\delta a^{2+\delta}$ carried by each interaction vertex in Eq.~(\ref{eq: SIGWs HI}) is exactly canceled.
We define $\vb{q}\equiv \vb{k}-\vb{p}$. For notational simplicity, explicit time arguments inside the integrals are omitted: all fields under a time integral are evaluated at the corresponding integration time.

The counterterm contribution is
\begin{align}\nonumber
	\mathcal{P}_h^{\rm cts}(\tau,k) & = \frac{k^{3+\delta}\mu^\delta}{4\pi^2} 8 \Im \Bigg\{ h^2_k \int_{-\infty_-}^\tau\dd \tau' \, a^{2+\delta} \Bigg\{  -\left[ C_1 \left( \dfrac{1}{2}\dot H + H^2\right) + \dfrac{\tilde m_\chi^2}{4}\right]  \left( h'^{*2}_{k} - k^2 h^{*2}_{k} \right) \\ \nonumber
	&- C_2 \dot H \left( h'^{*2}_{k} + \dfrac{k^2}{3}  h^{*2}_{k} \right) -C_3\left[\dfrac{k^2}{a^2}\left( k^2 h^{*2}_{k} -  h'^{*2}_{k}\right) + \dot H h'^{*2}_{k} \right] -C_4\left(\dfrac{2}{3}\dot H + H^2 \right) k^2 h^{*2}_{k}  \\
	&-  \delta \, \dot H \,  \left[   - \dfrac{C_1}{8} \left( h'^{*2}_{k} - k^2 h^{*2}_{k}\right) -\dfrac{C_3}{2} h'^{*2}_{k}  +\dfrac{C_4 - C_2}{9}k^2 h^{*2}_{k} \right] \Bigg\} \Bigg\}\,, \label{eq: SIGWs P cts}
\end{align} 
with the same convention for time arguments as above. As discussed in the previous section, the $\order{\delta}$ terms in the counterterm contribution must be retained because they can leave finite remnants after multiplying the poles in the counterterm coefficients.

To perform the loop integrals, it is convenient to trade the angular variable $\theta$ for the momentum $q\equiv \abs{\vb{k}-\vb{p}}$. The volume element becomes (see e.g.~\cite{Ballesteros:2024cef})
\begin{equation} \label{eq: SIGWs change to p q}
	\dd^{3+\delta}\vb{p} = p^{2+\delta} \left( \sin \theta \right)^{1+\delta} \dd p \, \dd \theta \, \dd \phi = p^\delta \sin^\delta\theta \, \frac{p\, q}{k}\, \dd p\, \dd q \, \dd \phi\,,
\end{equation}
with $q\in[\abs{k-p},\, k+p]$ and
\begin{equation}
	\sin^2 \theta = \dfrac{(k+p+q) (-k+p+q) (k-p+q) (k+p-q)}{4 k^2p^2} \,.
\end{equation}
The integration over the additional angular variables in $3+\delta$ spatial dimensions only produces an overall scheme-dependent normalization, as does the extra-dimensional Fourier factor $(2\pi)^{-\delta}$. Since both factors reduce to unity in three spatial dimensions, we set them to one without loss of generality. 
As explained in Section~\ref{sec: Dim Reg}, it is sufficient to expand the integrand to first order in $\delta$, while keeping $p^\delta$ unexpanded. Although the expansion of $\sin^\delta\theta$ generates logarithmic terms, these do not produce $\log p$ contributions in the large-$p$ limit, and therefore do not affect the UV analysis described in Section~\ref{sec: Dim Reg}, see \cite{Ballesteros:2024cef} for the details.

\subsubsection{Ultraviolet contribution and renormalized spectrum} \label{sec: SIGWs UV}

We now isolate the UV part of the scalar-induced tensor spectrum and renormalize it without specifying either the spectator potential $V(\overline{\chi})$ or the background evolution $H(\tau)$. We will show that, in dimensional regularization, the UV divergences of the tensor spectrum are absorbed by the covariant counterterms in Eq.~(\ref{eq:Cts action general}).

Following Section~\ref{sec: Dim Reg}, each loop contribution is split into an infrared and an ultraviolet part,
\begin{equation}
	\mathcal{P}_h \equiv \,^{\rm (IR)}\mathcal{P}_h + \,^{\rm (UV)}\mathcal{P}_h\,,
\end{equation}
corresponding to $p\in[0,L]$ and $p\in[L,\infty)$, respectively. In the present setup the IR part is finite and can be evaluated directly in $3$ spatial dimensions, whereas the UV part must be treated in $3+\delta$ spatial dimensions. Whenever a loop contribution contains a pole in $\delta$, we write
\begin{equation}
	\mathcal{P}_h \equiv \dfrac{1}{\delta}\,_{d}\mathcal{P}_h + \,_{f}\mathcal{P}_h\,,
\end{equation}
where $\,_{d} \mathcal{P}_h$ denotes the coefficient of the $1/\delta$ pole and $\,_{f}\mathcal{P}_h$ the finite part. We also define the total scalar loop contribution as
\begin{equation} \label{eq: SIGWs def P loop tot}
	\mathcal{P}_h^{\rm loop} = \mathcal{P}_h^{\rm q} + \mathcal{P}_h^{\rm c,1} + \mathcal{P}_h^{\rm c,2} \,,
\end{equation}
see Eqs.\ (\ref{eq: SIGWs P q}) and (\ref{eq: SIGWs P c}).

We begin by analyzing the UV region of the different one-loop diagrams contributing to the tensor spectrum.
For the quartic diagram in Eq.~(\ref{eq: SIGWs P q}), the large-$p$ expansion of the scalar modes gives
\begin{equation} \label{eq: Ph q UV}
	^{\rm (UV)}\mathcal{P}_h^{\rm q}(\tau,k) = \frac{k^{3+\delta}}{4\pi^2} \int _{p>L} \frac{\dd^{3+\delta} \vb{p}}{(2\pi)^3} \left(1+\dfrac{3\delta}{4} \right)  p^2 \sin^2\theta  \Im{h_{k}^2 \int_{-\infty_-}^\tau \dd \tau' \, h_{k}^{*2} \dfrac{1}{2W_p^\chi}}\,.
\end{equation}
Here $W_p^\chi(\tau')$ denotes the WKB frequency of the scalar modes in Eq.~(\ref{eq: SIGWs WKB}), whose large-momentum expansion is given in Eq.~(\ref{eq: WKB frec}). The UV momentum integrand is then reduced to an asymptotic expansion in powers of $p$. The explicit expression is reported in Appendix~\ref{app: formulas} and collected in Eq.~(\ref{eq: Ph q UV sol}).

The two diagrams with cubic insertions are treated using the general UV formulas for two-vertex in-in integrals, Eqs.~(\ref{eq: Dim Reg Ints Tempo 2 Vert I})--(\ref{eq: Dim Reg Ints Tempo 2 Vert II}). They give
\begin{align} 
	\nonumber
	^{\rm (UV)}\mathcal{P}_h^{\rm c,1}(\tau,k)  =& \frac{k^{3+\delta}}{4\pi^2} \int_{p>L}\frac{\dd^{3+\delta}\vb{p}}{(2\pi)^3} \left(\frac{1}{2}+\dfrac{\delta}{4} \right)  p^4 \sin^4\theta \sum_{{n,m} = 0}^{{\infty}} \dfrac{{i^{n-m}}}{(p+q)^{2+n+m}}  \\
	& \times \abs{h_k}^2 \partial^m_{\tau}\, \left(e^{i(p+q)\tau}h_k \chi_{c,p} \chi_{c,q}\right) \, \partial^n_{\tau}\left(e^{-i(p+q)\tau}h^*_k  \chi_{c,p}^*  \chi_{c,q}^*\right)  \,, \label{eq: Ph c1 UV} \\ \nonumber
	^{\rm (UV)}\mathcal{P}_h^{\rm c,2}(\tau,k)  =& \frac{k^{3+\delta}}{4\pi^2} \int_{p>L}\frac{\dd^{3+\delta}\vb{p}}{(2\pi)^3} \left(\frac{1}{2} +\dfrac{\delta}{4} \right)  p^4 \sin^4\theta \sum_{n = 0}^\infty \dfrac{{i^{n-1}}}{(p+q)^{1+n}} \\
	&\times \Re \bigg\{ -2h_k^2 \int_{-\infty_-}^\tau \mkern-10mu \dd \tau' \, \left(e^{i(p+q)\tau'}h^*_k \chi_{c,p} \chi_{c,q}\right)  \partial^n_{\tau'} \left(e^{-i(p+q)\tau'}h_k^* \chi_{c,p}^* \chi_{c,q}^*\right) \bigg\} \,. \label{eq: Ph c2 UV}
\end{align}
As discussed in Section~\ref{sec: Dim Reg}, although the expressions contain infinite sums, only terms scaling at least as $1/p$ in the loop momentum contribute to the UV part. Terms suppressed as $\order{1/p^2}$ generate $\order{1/L}$ corrections, which vanish after combining the IR and UV regions and taking $L\to\infty$, as in Eq.~(\ref{eq: Dim Reg First Paper}). 

A useful simplification occurs once the WKB form of the scalar modes in Eq.~(\ref{eq: SIGWs WKB}) is inserted. At first sight, the phases $\exp\left[\pm i\int^\tau \dd\tau' W_k^\chi(\tau')\right]$ seem to require the explicit \emph{model-dependent} evaluation of integrals over the WKB frequency in order to characterize the scalar modes in the large-momentum expansion. However, after the time-derivative operators generated by the UV expansion have acted on the modes in Eqs.~(\ref{eq: Ph c1 UV})--(\ref{eq: Ph c2 UV}), each scalar mode is paired with its complex conjugate evaluated at the same time. These phases therefore cancel exactly, so the model-dependent phase integrals are not needed to characterize the UV contribution. The derivatives of the phases only generate (non-integrated) factors of $W_k^\chi$ and its time derivatives, which can be expanded directly at large momentum using Eq.~(\ref{eq: WKB frec}). The resulting loop integrals are reported in Eqs.~(\ref{eq: Ph c1 UV sol}) and (\ref{eq: Ph c2 UV sol}).

After integrating by parts and using the free tensor equation of motion in $3+\delta$ spatial dimensions,
\begin{equation}
	h''_k + (2+\delta)aH \, h'_k + k^2 h_k = 0\,,
\end{equation}
the divergent part of the full UV loop contribution is
\begin{equation} \label{eq: SIGWs Loop div}
	\prescript{\rm (UV)}{d}{\mathcal{P}}_h^{\rm loop}(\tau,k) = \frac{k^3}{960 \pi ^4 } \Im \Bigg\{ h_k^2 \int_{-\infty_-}^\tau \dd \tau'  a^2 \, \bigg[ 2 \dot H  k^2 h_k^{*2} + (h'^{*2}_k-k^2 h_k^{*2}) \left(6 \dot H+9 H^2-5 m_\chi^2 \right)\bigg]\Bigg\} \,.
\end{equation}
All background quantities and mode functions inside time integrals are evaluated at the integration time $\tau'$. The corresponding finite UV contribution $\prescript{\rm (UV)}{f}{\mathcal{P}}_h^{\rm loop}$ is given in Eq.~(\ref{eq: Ph loop UV f sol}).

The divergence in Eq.~(\ref{eq: SIGWs Loop div}) has precisely the same tensor structures as the counterterm contribution in Eq.~(\ref{eq: SIGWs P cts}). Requiring the full spectrum to be finite fixes the pole parts of the counterterm coefficients,
\begin{align}
	\nonumber
	&C_1 = \dfrac{3}{640 \pi^2 \, \delta } + C_{f,1} \,, \quad C_2 = \dfrac{1}{1280 \pi^2 \, \delta } + C_{f,2} \,,\\
	&  C_3 = C_{f,3} \,,\quad  C_4 = C_{f,4} \quad {\rm and} \quad \tilde m_\chi^2 = - \dfrac{m_\chi^2}{96\pi^2 \, \delta} + \tilde m_{f,\chi}^2 \,. \label{eq: SIGW coef cts}
\end{align}
The finite parts $C_{f,i}$ and $\tilde m_{f,\chi}^2$ remain free parameters to be fixed by renormalization conditions, or equivalently by matching the observable to measurements. Notice that neither $R_{\mu\nu\rho\sigma}^2$ nor $\Box R$ is required to cancel the UV pole of the tensor spectrum, as reflected by the fact that $C_3$ and $C_4$ contain no pole part; see Appendix~\ref{app: Cts action SIGWS}. This is noteworthy because $\Box R$ is a total derivative and, in four dimensions, the Gauss--Bonnet combination
\begin{equation}
	R_{\mu\nu\rho\sigma}^2 - 4 R_{\mu\nu}^2 + R^2
\end{equation}
is topological, so that $R_{\mu\nu\rho\sigma}^2$ can be traded for a combination of $R_{\mu\nu}^2$ and $R^2$ up to a boundary term.

This observation requires some care in the in-in formalism, where temporal boundary terms can contribute to equal-time expectation values. Nevertheless, the covariant counterterm action is fixed independently of the observable and of the background around which it is expanded. The fact that no $\Box R$ or $R_{\mu\nu\rho\sigma}^2$ counterterm is required here is consistent with the existence of observables without temporal boundaries, for which pure boundary operators cannot be needed to cancel UV divergences. This does not forbid the use of such counterterms; it simply emphasizes that they are not required by observables without boundaries.

The renormalized one-loop scalar-induced tensor spectrum can be written as
\begin{align} \label{eq: SIWGs Final Spectrum}
	\nonumber
	&\mathcal{P}_h^{\rm total}(\tau,k) \equiv \mathcal{P}_h^{\rm loop}(\tau,k) + \mathcal{P}_h^{\rm cts}(\tau,k) = \lim_{L\to \infty}\Bigg[ \,^{\rm (IR)} \mathcal{P}_h^{\rm loop}(\tau,k)  - \frac{k^3 }{960 \pi ^4} \, L \, \Im{h_kh'^*_k}^2 \\ \nonumber
	& \qquad +  \frac{k^3}{960 \pi ^4 } \Im \Bigg\{ h_k^2 \int_{-\infty_-}^\tau \dd \tau'  a^2 \, \bigg[ 2 \dot H  k^2 h_k^{*2} + (h'^{*2}_k-k^2 h_k^{*2}) \left(6 \dot H+9 H^2-5 m_\chi^2 \right)\bigg] \log \dfrac{L}{a \mu}\Bigg\}  \\ \nonumber
	& \qquad - \dfrac{k^3}{120 \pi^4}  \Im \Bigg\{ h_k^2 \int_{-\infty_-}^\tau \dd \tau' \bigg[ L^4 h_k^{*2}  +\frac{L^2}{56}  \left(h_k^{*2} \left(28 a^2 \left(\dot H +2 H^2-m^2_\chi \right)+11 k^2\right)-7 h_k'^{*2}\right) \bigg]  \Bigg\}\Bigg]\\ \nonumber
	& \qquad + \dfrac{k^3}{20160 \pi^4} \Im\Bigg\{ h_k^2 \int_{-\infty_-}^\tau \dd \tau' a^2 \Bigg[ h'^{*2}_k \bigg[ -\frac{C_{f,2} k^2}{a^2}-63 C_{f,1} H^2-C_{f,3} \, \dot H+15 \tilde m_{f,\chi}^2  \bigg] \\ \nonumber
	& \hspace{5.2cm} + h^{*2}_k \bigg[ \frac{(2+C_{f,2}) k^4}{a^2}+21 a^2 \left(\dddot H +3 \left(3 \ddot H  H+6 \dot H  H^2+\dot H ^2\right)\right) -15 k^2 \tilde m_{f,\chi}^2 \\ \nonumber
	& \hspace{5.2cm} +6 m_\chi^2  \left(7 a^2 \left(2 \dot H +3 H^2\right)+5 k^2\right) -63 a^2 m_\chi^4 -84 a H (m_\chi^2)' -21 (m_\chi^2)''  \\
	& \hspace{5.2cm} + \frac{k^2}{6} \left( \dot H  (111-2  C_{f,2} -2 C_{f,3}+84 C_{f,4})+126 C_{f,4} H^2\right) \bigg] \Bigg]  \Bigg\} \,.
\end{align}
In writing this expression, scheme-dependent constants, which carry no intrinsic information, have already been absorbed into the finite counterterm parameters, chosen to simplify the remaining finite contribution.
We emphasize that the only assumption needed in the derivation of Eq.~(\ref{eq: SIWGs Final Spectrum}) is that the loop modes admit the asymptotic past positive-frequency behavior in Eq.~(\ref{eq: SIGWs asymp exp}).

Although Eq.~(\ref{eq: SIWGs Final Spectrum}) is necessarily lengthy, its structure is simple. The terms inside the $\lim_{L\to\infty}$ combine the IR loop integral over $p\in[0,L]$ with the explicit $L$-dependent UV contributions, which cancel the artificial dependence on the splitting scale $L$. The remaining $L$-independent part contains the finite UV contribution together with the finite part of the counterterms.
The latter includes both the free finite counterterm coefficients and the finite remnants fixed by the pole counterterms.

One such remnant comes from the overall extra-dimensional factor in the counterterm integrand, cf.~Eq.~(\ref{eq: SIGWs P cts}),
\begin{equation} \label{eq: SIGWs a to the delta diss}
	(k\mu a(\tau'))^\delta = 1 + \delta \, \log(k\mu a(\tau')) + \order{\delta^2} \,.
\end{equation}
At leading order in the UV pole,
\begin{equation} \label{eq: SIGWs a to the delta diss 2}
	\mathcal{P}_h^{\rm cts}(\tau,k) = -\dfrac{1}{\delta} \prescript{\rm (UV)}{d}{\mathcal{P}}_h^{\rm loop}(\tau,k) \,.
\end{equation}
The $\order{\delta}$ term in Eq.~(\ref{eq: SIGWs a to the delta diss}) combines with the $1/\delta$ pole in the counterterm coefficients of Eq.~(\ref{eq: SIGWs a to the delta diss 2}), leaving a finite remainder that yields the scale-invariant logarithm $\log\left(L/(\mu a(\tau'))\right)$ appearing in Eq.~(\ref{eq: SIWGs Final Spectrum}).

The cancellation of the UV pole by the restricted set of covariant counterterms is a remarkable consistency check. The counterterms are not an arbitrary local quadratic basis for $h_{ij}$: general covariance fixes their tensor structures and their time dependence through background quantities. In particular, no tensor mass counterterm $h_{ij}^2$ is available. The fact that the divergence Eq.~(\ref{eq: SIGWs Loop div}) is absorbed by the counterterms descending from Eq.~(\ref{eq:Cts action general}) therefore confirms that the UV part of the scalar loop is compatible with covariant EFT renormalization on a time-dependent background.

Since Eq.~(\ref{eq: SIWGs Final Spectrum}) is finite, the renormalized spectrum can be evaluated directly in three spatial dimensions, even though its derivation required working in $3+\delta$ dimensions. Because no specific choice of $V(\overline{\chi})$ or $H(\tau)$ has been made, the result is particularly well suited for numerical applications: the UV subtraction has already been performed, and the expression requires no further renormalization.

It is instructive to contrast the renormalization procedure based on dimensional regularization with what happens in a hard-cutoff scheme. For simplicity, let us consider a comoving cutoff, even though a physical cutoff is more appropriate in cosmological backgrounds, as explained in \cite{Senatore:2009cf} (see also \cite{Hauser:2026jnp}). The relevant divergences are already visible in Eq.~(\ref{eq: SIWGs Final Spectrum}). For instance, the quartic divergence, proportional to $\Lambda^4$, is obtained from the $L^4$ term with the opposite sign, since that term is precisely the UV subtraction that cancels the quartic divergence of the loop. Its structure coincides with that generated by a tree-level insertion of a tensor mass counterterm. Removing this divergence would therefore require introducing a counterterm proportional to $h_{ij}^2$, which is absent from the covariant basis built out of curvature invariants. This reflects the fact that a hard cutoff breaks diffeomorphism invariance and forces one to enlarge the counterterm basis; see \cite{Ballesteros:2024cef} for further details, and \cite{Fang:2026off} for a related discussion of Ward-identity-based renormalization conditions (relevant for symmetry-breaking regulators).

\subsection{A particular case: displaced-mass in de Sitter} \label{sec: SIGWs dS}

We now apply the renormalized result of the previous section to a simple de Sitter example that can be treated analytically and that nevertheless contains a non-trivial comoving scale.\footnote{See \cite{Campos:1994xx,Frob:2012ui} for related calculations of the tensor spectrum induced by a scalar field conformally coupled to gravity. In that case, a conformal transformation maps the relevant part of the computation to an equivalent one on a flat background, where the loop integrals can be evaluated using standard dimensional-regularization techniques.}
We take $H$ to be constant and choose the spectator mass profile
\begin{equation}
	m_\chi^2(\tau) = 2 H^2\left( 1- \dfrac{\tau^2}{(\tau - \tau_*)^2}\right)  \,,
\end{equation}
with $\tau_*>0$. Since the physical inflationary range is $\tau<0$, the pole at $\tau=\tau_*$ lies outside the integration domain. In three spatial dimensions this profile gives
\begin{equation}
	m_{{\rm eff},\chi}^{2}(\tau) =  - 2 H^2 a^2\eval_{\tau-\tau_*} \,.
\end{equation}
Thus, the spectator fluctuation has the same effective mass as a massless scalar in de Sitter, but with the time coordinate displaced by $\tau_*$.

The scale $\tau_*$ makes the late-time spectrum depend on the scale-invariant combination
\begin{equation}
	\kappa \equiv k \tau_* >0\,.
\end{equation}
Any residual dependence on $\kappa$ that cannot be reproduced by the finite parts of local counterterms represents a distinguishable loop contribution. The limit $\tau_*\to0$, or equivalently $\kappa\to0$ at fixed $k$, smoothly recovers the massless de Sitter case studied in \cite{Ballesteros:2024cef}. In this limit no additional scale is present and, since no late-time divergence arises as $\tau\to0$, the result remains scale invariant and is therefore scheme dependent.

The mode functions of the canonically normalized fields can be obtained directly in three spatial dimensions from Eqs.~(\ref{eq: SIGWs eom scalar})--(\ref{eq: SIGWs eom tensor}). Since Eq.~(\ref{eq: SIWGs Final Spectrum}) is already renormalized, we emphasize that no further $3+\delta$ treatment is required at this stage. Choosing the standard positive-frequency normalization in the asymptotic past, we find
\begin{equation}
	\chi_{c,k}(\tau) =  \frac{e^{-i k \tau }}{\sqrt{2 k}}  \left(1-\frac{i}{k (\tau -\tau_*)}\right)\,, \quad h_{c,k}(\tau) = \dfrac{e^{-i k \tau}}{\sqrt{2k}} \left( 1 - \frac{i}{k \tau}\right)  \,.
\end{equation}
With these mode functions, the time and momentum integrals in Eq.~(\ref{eq: SIWGs Final Spectrum}) can be performed analytically. We first carry out the time integrals, with the usual $i\epsilon$ prescription ensuring convergence in the asymptotic past. The logarithmic time integrals appearing in Eq.~(\ref{eq: SIWGs Final Spectrum}) require some care; their analytic treatment is presented in Appendix~\ref{app: Log time int}. At this stage, the only remaining loop integral is the finite IR contribution, $\prescript{\rm (IR)}{}{\mathcal{P}}_h^{\rm loop}$, with loop momentum restricted to $p\in[0,L]$, which, as mentioned, can also be evaluated in closed form.

We now give the late-time limit of the renormalized tensor spectrum. This limit is finite, with no residual $\log(-k\tau)$ dependence, and its momentum dependence is therefore a function only of $\kappa$. The finite counterterm freedom that survives at late times reduces to two contributions parametrized below by $\alpha$ and $\overline m_{f,\chi}^2$. The scheme-dependence on $\mu/H$ has been absorbed into these finite parameters.\footnote{For completeness, the finite redefinitions from Eq.~(\ref{eq: SIWGs Final Spectrum}) used to write Eq.~(\ref{eq: Ph dismass LT}) are
	\begin{equation}
		C_{f,1} \equiv 6 \left( \alpha + \log\frac{2\mu}{H} + \gamma_E\right) +\frac{C_{f,2}}{21}+C_{f,4}+\frac{377}{42} \,, \quad \tilde m_{f,\chi}^2 \equiv \overline{m}_{f,\chi}^2 - 7 m_\chi^2 \left( \log\frac{2\mu}{H}+\gamma_E  +\frac{3}{10}\right) \,.
\end{equation}} We find:
\begin{align} \label{eq: Ph dismass LT}
	\nonumber
	 &\mathcal{P}_h^{\rm total} (\tau \to 0,k) =\left(\dfrac{H}{\pi M_P} \right)^4  \Re\Bigg\lbrace\frac{1}{240} \left(9 \alpha -10 \pi  \kappa +\frac{5 \pi }{\kappa }\right)+\frac{1}{12} \left(\log^2(2\kappa)+(2 \gamma_E -i \pi ) \log(2\kappa) \right) e^{-2 i \kappa } \kappa ^2\\ \nonumber
	 &\quad-\frac{1}{12} ({\rm Ei}(-2i\kappa)+i \pi ) ({\rm Ei}(2i\kappa)-i \pi ) \kappa ^2 +\frac{i}{3}  \, _3F_3(1,1,1;2,2,2;2 i \kappa ) e^{-2 i \kappa } \kappa ^3\\
	 &\quad+\frac{e^{-2 i \kappa } A_1(\kappa)}{144 \kappa}+i\frac{ e^{-2 i \kappa } {\rm Ei}(2i\kappa)  A_2(\kappa)}{24 \kappa }\Bigg\rbrace +\dfrac{k^3}{1344 \pi^4} \Im\Bigg\{ h_k^2 \int_{-\infty_-}^\tau \dd \tau' a^2\overline m_{f,\chi}^2 \Bigg[ h'^{*2}_k  - k^2 h^{*2}_k \Bigg]  \Bigg\}\,,
\end{align} 
where 
\begin{align}
	A_1(\kappa) =12 \gamma_E ^2 \kappa ^3+11 \pi ^2 \kappa ^3+6 \pi  (-1-2 i \kappa  (1+\kappa  (\gamma_E  \kappa +2 i)))\,, \quad A_2(\kappa) = -1+2 \kappa  (\kappa  (\pi  \kappa +2)-i)\,.
\end{align}
Here $\gamma_E$ is the Euler--Mascheroni constant and ${}_pF_q$ denotes the generalized hypergeometric function.

\begin{figure}[t!]
	\centering
	\includegraphics[width=0.6\textwidth]{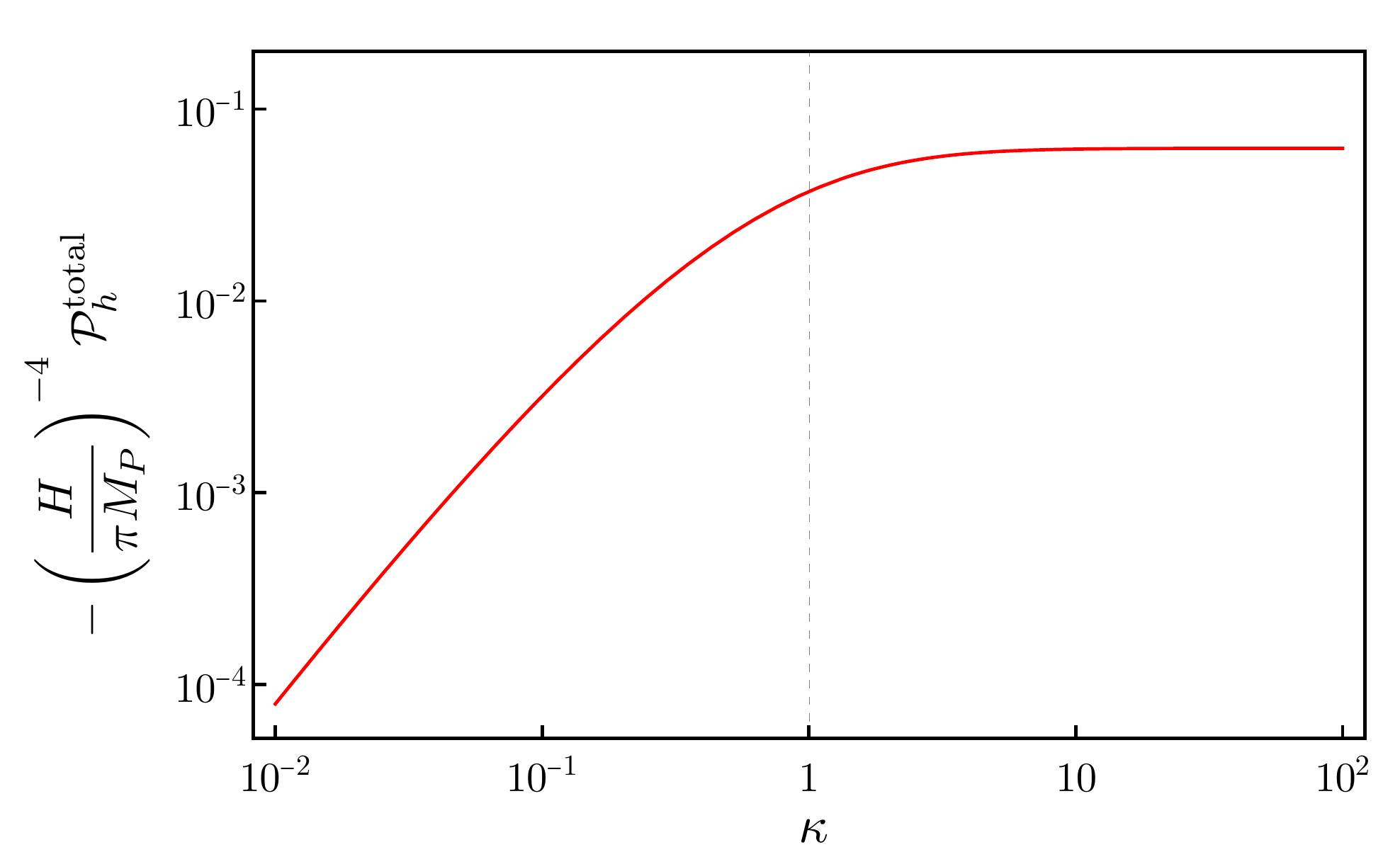}
	\caption{\small{\it Negative of the renormalized one-loop tensor spectrum in Eq.~(\ref{eq: Ph dismass LT}), normalized by $(H/(\pi M_P))^4$, as a function of $\kappa$. We set $\overline m^2_{f,\chi}=0$ and choose $\alpha$ so that the one-loop contribution vanishes exactly in the limit $\kappa\to0$.}}
	\label{fig}
\end{figure}

The last term of Eq.~(\ref{eq: Ph dismass LT}) comes from the finite part of the covariant counterterm $\tilde V''R$. We keep this contribution explicit in Eq.~(\ref{eq: Ph dismass LT}) in order to display the full counterterm freedom. A natural choice would be to take it proportional to $m_\chi^2$, namely to assign it the same time dependence multiplied by a finite constant coefficient, $\overline m_{f,\chi}^2\propto m_\chi^2$, in analogy with Eq.~(\ref{eq: SIGW coef cts}). However, the EFT allows for a more general finite profile. In the discussion below, we set $\overline m_{f,\chi}^2=0$ in order to isolate the intrinsic dependence on $\kappa$, see Fig.~\ref{fig}.

The two limiting regimes are
\begin{align} \label{eq: SIGWs Ph LT IR}
	&\mathcal{P}_h^{\rm total} (\tau \to 0,k \to 0) =  \left(\dfrac{H}{\pi M_P} \right)^4 \left( \frac{1}{12} + \frac{3 \alpha }{80}+\frac{1}{216} \kappa ^2 \left(48 \log (2 \kappa )+3 \pi ^2+48 \gamma_E -40\right) +\order{\kappa^3}\right) \,,\\
	&\mathcal{P}_h^{\rm total} (\tau \to 0,k \to \infty) = \left(\dfrac{H}{\pi M_P} \right)^4 \left( \frac{1}{48} + \frac{3 \alpha }{80}+\frac{3}{64 \kappa ^2}+\order{\kappa^{-3}}\right) \,.
\end{align}
The spectrum is finite in both limits. In the IR, $\kappa\ll1$, the leading contribution is scale invariant and can be absorbed into the finite local parameter $\alpha$, thereby recovering the known de Sitter result for a massless scalar field \cite{Ballesteros:2024cef}.\footnote{We emphasize that scale invariance does not imply an identically vanishing one-loop contribution, as found in \cite{Ema:2025ftj}, and also in the scalar sector in \cite{Fumagalli:2023zzl, Fumagalli:2024jzz, Tada:2023rgp, Inomata:2024lud, Inomata:2025pqa, Ema:2026dop}. See \cite{Hauser:2026jnp} for a recent discussion of the origin of this putative cancellation. In the present case, its failure is already manifest from Eq.~(\ref{eq: SIGWs Ph LT IR}).} Physically, these modes are already far outside the horizon when the displaced scale $\tau_*$ becomes relevant, and therefore cannot resolve the time displacement. The first non-trivial imprint of the displaced mass appears in the subleading $\kappa^2\log\kappa$ dependence. In the opposite regime, $\kappa\gg1$, the spectrum again approaches a constant, up to power-suppressed corrections. The non-trivial dependence on $\kappa$ at intermediate scales, and in particular the terms that cannot be mimicked by the available finite counterterms, constitute the distinguishable part of the loop correction.

We emphasize that Eq.~(\ref{eq: Ph dismass LT}) has a finite late-time limit: no term of the form $\log(-k\tau)$ survives as $\tau\to0$. Such late-time growth would signal an effective tensor mass generated by the loop, whose effects are studied in \cite{Ballesteros:2024cef} in de Sitter. Its absence is therefore an important consistency check of the freeze-out of tensor modes at loop level, in close parallel with the standard arguments for the scalar metric fluctuation $\zeta$, see e.g.~\cite{Pimentel:2012tw,Assassi:2012et,Senatore:2012ya}. It also illustrates the power of dimensional regularization: regulators that break diffeomorphism invariance can produce spurious mass-like terms, which must then be canceled by non-covariant counterterms \cite{Ballesteros:2024cef}.

\section{One-loop scalar spectrum} \label{sec: USR}

We now apply the renormalization framework developed above to the scalar power spectrum generated by scalar self-interactions coming from a potential. We consider
\begin{equation}
	S = \int \dd^{4+\delta} x \, \mu^\delta  \,\sqrt{-g} \left(- \dfrac{1}{2} g^{\mu \nu} \partial_\mu \phi \partial_\nu \phi - V(\phi) \right) 
\end{equation}
in $3+\delta$ spatial dimensions. Our goal is to isolate the UV structure of the one-loop correction and to show how the corresponding divergences are absorbed by local counterterms, leading to a finite expression that can be used without specifying the detailed background evolution.

The class of scenarios considered here is defined by the regime in which the scalar dynamics relevant for the loop is accurately captured by the potential, while the interactions induced by the metric fluctuations, namely the lapse and the shift, are subleading. This is relevant, for instance, in single-field inflation with a transient violation of slow-roll. In ultra slow-roll-like phases one may have $\epsilon \equiv - H'/(aH^2) \ll 1$, while higher slow-roll parameters are $\order{1}$, so that derivatives of the potential generate sizable scalar vertices. In this limit the lapse- and shift-mediated interactions are subleading for the scalar correlator considered below \cite{Maldacena:2002vr}; see also \cite{Wang:2013zva,Ballesteros:2024zdp}. 
In this class of scenarios, \cite{Kristiano:2025ajj} showed that tadpole renormalization fixes the counterterm structure required to cancel the UV divergences of the scalar spectrum. The same tadpole-renormalized setup has recently been used in \cite{Inomata:2026csq} to argue that, for modes already on superhorizon scales well before the transient violation of slow-roll, the scalar spectrum remains conserved; see also \cite{Inomata:2025pqa}. Using the method of Section~\ref{sec: Dim Reg}, we explicitly recover the UV cancellation in the power spectrum implied by tadpole renormalization found in \cite{Kristiano:2025ajj}, and derive for the first time the final renormalized one-loop scalar spectrum in closed form, while keeping the background evolution arbitrary.

Once loop corrections are included, the action above has to be supplemented by local counterterms. For the potential interactions considered here, these counterterms can be organized in a particularly simple way: a field renormalization together with a counterterm potential, 
\begin{equation}
	\phi\to (1+Z_\phi/2) \phi \quad {\rm and} \quad V(\phi) \to V + \tilde{V} \eval_{(1+Z_\phi/2) \phi }\,,
\end{equation}
where $Z_\phi$ is a dimensionless constant.
We will keep only the terms linear in the counterterm insertions. Products such as $Z_\phi^2$ or $Z_\phi \tilde V$ are therefore consistently neglected, since they correspond to higher-order counterterm contributions. Setting the metric to its background value, the action becomes 
\begin{equation} 
	S = \int  \dd \tau \, \dd^{3+\delta} \vb{x} \, \mu^\delta  \, a^{4+\delta} \left[  \dfrac{(1+Z_\phi/2)^2}{2a^2}\left(  \phi'^2 - (\partial_i \phi)^2\right)  - \left( V+ \tilde{V} \right)\eval_{(1+Z_\phi/2) \phi }  \right] \,.
\end{equation}
We work in the $\delta\phi$-gauge and decompose the scalar field into its background value and fluctuation, 
\begin{equation}
	\phi(x) = \phi_0(t) + \delta \phi(x) \,.
\end{equation} 
We use the notation
\begin{equation}
	V(\phi) = \sum_n \dfrac{V_n(\phi_0)}{n!} \delta \phi^n \,, \quad {\rm where} \quad V_n(\phi_0) \equiv \dfrac{\partial^n V(\phi)}{\partial \phi^n} \eval_{\phi_0} \,,
\end{equation}
and analogously for $\tilde V_n(\phi_0)$. Along the background trajectory this implies 
\begin{equation}
	V_n(\phi_0) = \dfrac{V_{n-1}'(\phi_0)}{\phi_0'} = \order{\epsilon^{(2-n)/2}} \,.
\end{equation}
This is the power counting appropriate to the regime considered here: $\epsilon \ll 1$, suppressed with respect to the remaining slow-roll parameters. If the latter assumption is relaxed, the potential-induced vertices are no longer parametrically dominant, and the $\epsilon$-suppressed interactions generated by the lapse and shift must be retained as well.

Expanding the action in powers of $\delta\phi$, the free quadratic Lagrangian density $\left( L = \int \dd^{3+\delta} \vb{x} \, \mathcal{L}\right) $ is
\begin{equation}
	\mathcal{L}_0 =  \dfrac{\mu^\delta a^{2+\delta}}{2} \left( \delta \phi'^2 - (\partial_i \delta \phi)^2 - a^2 \delta \phi^2 V_2(\phi_0)\right) \,.
\end{equation}
This defines the free Hamiltonian and therefore the dynamics of the fields in the interaction picture. The interaction Lagrangian is obtained by subtracting $\mathcal{L}_0$ from the full Lagrangian. In the presence of velocity-dependent interactions, the corresponding interaction Hamiltonian density ($H = \int \dd^{3+\delta} \vb{x} \, \mathcal{H}$) is not simply $-\mathcal{L}_I$. Rather, for a quadratic kinetic term normalized as above,
\begin{equation}
	\mathcal{H}_I = - \mathcal{L}_I + \dfrac{1}{2 \mu^\delta a^{2+\delta}} \left(\dfrac{\partial \mathcal{L}_I}{\partial \delta \phi'} \right)^2 + \order{ \mathcal{L}_I}^3 \,,
\end{equation}
see e.g.~\cite{Pimentel:2012tw,Ballesteros:2025nhz}. Applying this relation and keeping only the terms needed for the one-loop scalar spectrum, we obtain
\begin{align} \label{eq: USR HI}
	\nonumber
	H_I =\int \dd^{3+\delta} \vb{x} \, \mu^{\delta} \Bigg[&-\partial_\tau\left( a^{2+\delta} \phi_0' \delta \phi\right) + a^{4+\delta}\left( \delta \phi^3 \dfrac{V_3}{3!} + \delta \phi^4 \dfrac{V_4}{4!}\right) \\
	& + a^{4+\delta} \left( \delta \phi \overline V_1 + \delta \phi^2 \frac{\overline{V}_2}{2}\right)  -\dfrac{Z_\phi}{2} \left( \partial_\tau\left( a^{2+\delta} \delta \phi' \delta \phi\right) +  a^{4+\delta} \delta \phi^2  V_2 \right)  \Bigg]\,.
\end{align}
Here and in what follows, $V_n=V_n(\phi_0)$ and $\overline V_n=\overline V_n(\phi_0)$, with the dependence on the background omitted for brevity. The counterterm potential appearing in Eq.~(\ref{eq: USR HI}) has been reorganized into an effective potential $\overline V$, defined along the background trajectory by 
\begin{equation} \label{eq: defpot}
	\overline{V}_1 \equiv \tilde V_1 + \frac{Z_\phi}{2}(V_1 + \phi_0 V_2) \,, \quad \textrm{and thus} \quad \overline{V}_2 = \dfrac{\overline V_1'}{\phi_0'} = \tilde V_2 +Z_\phi V_2+ \frac{Z_\phi}{2} \phi_0 V_3 \,.
\end{equation}
In deriving Eq.~(\ref{eq: USR HI}) we have discarded purely background terms and spatial boundary terms, and we used the background equation
\begin{equation}
	\phi_0'' + (2+\delta) aH \phi_0' + a^2V_1 = 0\,,
\end{equation}
together with the free equation of motion for the fluctuation,
\begin{equation}
	\delta \phi'' + (2+\delta) a H \delta \phi' - \partial^2 \delta \phi + a^2V_2 \delta \phi = 0 \,.
\end{equation}
Using the latter is legitimate because the fields appearing in the interaction Hamiltonian obey the free dynamics in the interaction picture.

The definition in Eq.~(\ref{eq: defpot}) makes the renormalization structure transparent. The first line of Eq.~(\ref{eq: USR HI}) contains the interaction terms inherited from the original action, whereas the second line contains the counterterm insertions. The first boundary term, which is linear in the fluctuation, does not contribute to the one-point function considered below and will not play any further role \cite{Arroja:2011yj,Burrage:2011hd, Braglia:2024zsl}. In contrast, the boundary term multiplied by $Z_\phi$ contributes to the scalar spectrum. The counterterm potential $\overline V$ renormalizes the tadpole and, once $\overline V_1$ is fixed, also fixes the tower $\overline{V}_n = \overline{V}'_{n-1} / \phi_0'$; see \cite{Kristiano:2025ajj, Inomata:2025pqa,Inomata:2026csq}. 

We finally specify the free dynamics. The equation of motion for the canonically normalized field is
\begin{align}  \label{eq: USR eom}
	\delta \phi''_{c,k} +  \left( k^2 + m_{{\rm eff}}^{2}(\tau)\right) \delta \phi_{c,k}  = 0 \,, \quad \lim_{\tau \to - \infty}  \delta\phi_{c,k} \sim \dfrac{e^{-i k \tau}}{\sqrt{2k}} \quad {\rm with} \quad \delta \phi_{c,k} \equiv \mu^{\tfrac{\delta}{2}} a^{\tfrac{2+\delta}{2}} \delta \phi_k \,.
\end{align}
The effective mass is
\begin{equation} \label{eq: USR meff}
	m_{{\rm eff}}^{2} \equiv a^2\left[ V_2 - \dfrac{2+\delta}{2} \dot H - \dfrac{4+3\delta}{2} H^2  \right] +\order{\delta^2}\,.
\end{equation}
In the class of models considered here, $\order{\epsilon}$ corrections are neglected, so the term proportional to $\dot H$ is subleading. We also assume that the far past contains a regime in which $k^2 \gg m_{\rm eff}^2$, so that the plane-wave behavior in Eq.~(\ref{eq: USR eom}) is compatible with the free equation of motion and selects the positive-frequency solution.

As in the tensor calculation, the large-momentum behavior of the canonical mode is described by the WKB expansion 
\begin{equation} \label{eq: USR WKB}
	\delta \phi_{c,k}(\tau) =  \dfrac{ e^{-i \int^\tau_{\tau_*} W_k(\tau') \dd \tau'}}{\sqrt{2W_k(\tau)}}  \,,
\end{equation}
which selects the positive-frequency solution dictated by the asymptotic vacuum condition. We emphasize that this expansion relies on the hierarchy $\abs{\dd^n m_{\rm eff}^2(\tau)/\dd\tau^n}\ll k^{n+2}$. If $m_{\rm eff}^2$ or any of its derivatives undergoes an abrupt transition, this condition fails: some derivatives become formally singular and the UV expansion of the modes is no longer described by the simple WKB form above. Indeed, in such scenarios the UV expansion of the modes generally mixes positive- and negative-frequency phases of the solutions to the equations of motion. Abrupt transitions should therefore be regarded as idealized approximations; see e.g.~\cite{Ballesteros:2024zdp}. For smooth transitions, the WKB expansion is valid and, as we will show, it allows for a model-independent renormalization of the one-loop scalar spectrum.

\subsection{Tadpole renormalization} \label{sec: USR tadpole}

Before turning to the power spectrum, we must first renormalize the tadpole. At tree level, imposing the background equations ensures that the action for fluctuations starts at quadratic order.\footnote{Although the interaction Hamiltonian in Eq.~(\ref{eq: USR HI}) contains a linear term, this term is a total derivative in time. Since it contains no time derivatives of the scalar fluctuation, it does not contribute to the one-point function \cite{Arroja:2011yj, Burrage:2011hd, Braglia:2024zsl}.} At one loop, the same statement becomes the renormalization condition $\expval{\delta\phi}=0$, which prevents the zero mode of the scalar fluctuation from shifting the background evolution fixed by $H(\tau)$.\footnote{This tadpole renormalization is sometimes described as \emph{backreaction}; see e.g.~\cite{Inomata:2024lud}. However, we emphasize that the background is kept fixed by the prescribed $H(\tau)$.} This is enforced by the linear counterterm in Eq.~(\ref{eq: USR HI}) (see e.g.~\cite{Cheung:2007st,Pimentel:2012tw}).

Using Eq.~(\ref{eq: In-In short 1}), the one-loop contribution generated by the cubic interaction, together with the counterterm contribution arising from the linear vertex, both encoded in Eq.~(\ref{eq: USR HI}), gives
\begin{equation}
	\expval{\delta \phi(x)} = \begin{tikzpicture}[baseline={-2}]
		\draw (1,-0.3) -- (1,0.3);
		\draw[fill=white,cross] (1,0.3) circle (0.15);
	\end{tikzpicture} + \begin{tikzpicture}[baseline={-2}]
		\draw (2,-0.3) -- (2,0.15);
		\draw (2,0.45) circle (0.3);
	\end{tikzpicture}  = \int \dd^{3+\delta} \vb{k} \, \delta(\vb{k}) \, 2 \Im{ \delta \phi_k \int_{-\infty_-}^\tau \mkern-10mu \dd \tau' \mu^\delta a^{4+\delta} \delta\phi_k^* \left[ \overline V_1 + \dfrac{V_3}{2} \int \dfrac{\dd^{3+\delta} \vb{p}}{(2\pi)^{3+\delta}} \abs{\delta\phi_p}^2\right] } \,.
\end{equation}
All quantities inside the time integral are evaluated at $\tau'$; the external field $\delta\phi_k$ is evaluated at time $\tau$. Imposing a vanishing tadpole then fixes the counterterm coupling $\overline V_1$, including both its divergent and finite parts, as
\begin{equation}
	\overline V_1 = - \dfrac{V_3}{2} \int \dfrac{\dd^{3+\delta} \vb{p}}{(2\pi)^{3+\delta}} \abs{\delta\phi_p}^2 \,.
\end{equation}
As anticipated, once $\overline V_1$ is fixed, the whole tower $\overline V_{n>1}$ is determined recursively. In particular, for the one-loop scalar power spectrum we will need
\begin{equation} \label{eq: USR tilde V2}
	\overline V_2 = \dfrac{\overline V'_1}{\phi_0'} = - \dfrac{V_4}{2} \int \dfrac{\dd^{3+\delta} \vb{p}}{(2\pi)^{3 + \delta}} \abs{\delta\phi_p}^2 - \dfrac{V_3}{2 \phi_0'} \int \dfrac{\dd^{3+\delta} \vb{p}}{(2\pi)^{3+\delta}} \left( \abs{\delta\phi_p}^2\right) ' \,.
\end{equation}
This tadpole renormalization is model independent in the sense that it does not require specifying the detailed background dynamics.

Although the importance of tadpole renormalization for a consistent loop-level treatment of inflationary correlators was already emphasized in \cite{Pimentel:2012tw}, its relevance for this class of scenarios, and more specifically in the ultra slow-roll context, was pointed out in \cite{Ballesteros:2024zdp,Inomata:2024lud}. A fully consistent implementation was later carried out in \cite{Kristiano:2025ajj,Inomata:2025pqa,Inomata:2026csq}, and also in \cite{Fumagalli:2024jzz}, following the logic of \cite{Pimentel:2012tw}. 

\subsection{One-loop scalar power spectrum}

Once tadpole renormalization has been established, we can proceed to the one-loop scalar power spectrum. We define the dimensionless spectrum as
\begin{equation}
	\expval{\delta \phi(\tau,\vb{x}) \delta\phi(\tau,\vb{y})}_c = \int \frac{\dd^{3+\delta}\vb{k}}{(2\pi)^3} e^{i\vb{k}\cdot (\vb{x} - \vb{y})} \frac{2\pi^2}{k^{3+\delta}} \mathcal{P}_{\delta \phi}(\tau,k) \,,
\end{equation}
where we again work in the renormalization scheme in which the extra-dimensional contribution associated with the Fourier-space volume element is absorbed into the finite part of the counterterms.

At one loop, the scalar spectrum receives three contributions,
\begin{equation}
	\mathcal{P}_{\delta \phi}^{\rm total} \equiv \mathcal{P}_{\delta \phi}^{V_3} + \mathcal{P}_{\delta \phi}^{V_4} + \mathcal{P}_{\delta \phi}^{\rm cts} = \begin{tikzpicture}[baseline={-2}]
		\draw (0,0) -- (1.5,0);
		\draw (0.5+0.25,0.25) circle (0.25);
	\end{tikzpicture} + \begin{tikzpicture}[baseline={-2}]
		\draw (0,0) -- (0.5,0);
		\draw (0.5+0.25,0) circle (0.25);
		\draw (1,0) -- (1.5,0);
	\end{tikzpicture} +  \begin{tikzpicture}[baseline={-2}]
	\draw (0,0) -- (1.5,0);
	\draw[fill=white,cross] (0.75,0) circle (0.15);
	\end{tikzpicture} \,.
\end{equation}
The contribution associated with the quartic interaction follows from Eq.~(\ref{eq: In-In short 1}) and is given by
\begin{equation} 
	\mathcal{P}_{\delta\phi}^{V_4}  = \dfrac{k^{3+\delta}}{2 \pi^2} \int \dfrac{\dd^{3+\delta} \vb{p}}{(2\pi)^{3}} \, 2\Im{\delta\phi_k^2 \int_{-\infty_-}^\tau \mkern-10mu \dd \tau' a^{2} \, \delta\phi_k^{*2} \dfrac{V_4}{2}  \abs{\delta\phi_{c,p}}^2 } \,.
\end{equation}
The diagram with two insertions of the cubic interaction Hamiltonian is obtained from Eq.~(\ref{eq: In-In short 2}) and splits into two terms,
\begin{align} \label{eq: USR P cubic}
	&\mathcal{P}_{\delta\phi}^{V_3} \equiv \mathcal{P}_{\delta\phi}^{V_3,1} + \mathcal{P}_{\delta\phi}^{V_3,2}\,, \quad {\rm with}  \\
	&\mathcal{P}_{\delta\phi}^{V_3,1} = \frac{k^{3+\delta}}{2\pi^2} \int\frac{\dd^{3+\delta}\vb{p}}{(2\pi)^3}  \abs{\delta\phi_k}^2  \int_{-\infty_+}^\tau \mkern-10mu \dd \tau'\, a^2V_3 \, \delta\phi_k \delta\phi_{c,p} \delta\phi_{c,q} \int_{-\infty_-}^\tau \mkern-10mu \dd \tau''\, a^2 V_3\, \delta\phi^*_k  \delta\phi_{c,p}^*  \delta\phi_{c,q}^* \,, \\
	&\mathcal{P}_{\delta\phi}^{V_3,2}= \frac{k^{3+\delta}}{2\pi^2} \int\frac{\dd^{3+\delta}\vb{p}}{(2\pi)^3}  \Re \bigg\{ -2\delta\phi_k^2 \int_{-\infty_-}^\tau \mkern-10mu \dd \tau' \, a^2 V_3\,  \delta\phi^*_k \delta\phi_{c,p} \delta\phi_{c,q}  \int_{-\infty_-}^{\tau'} \mkern-10mu \dd \tau'' \, a^2 V_3\,  \delta\phi_k^* \delta\phi_{c,p}^* \delta\phi_{c,q}^* \bigg\}\,, 
\end{align}
where $q \equiv \abs{\vb{k}-\vb{p}}$. In both loop diagrams, $\mathcal{P}_{\delta\phi}^{V_3}$ and $\mathcal{P}_{\delta\phi}^{V_4}$, we write the loop modes in terms of the canonically normalized field. In this way, the factors $\mu^\delta a^\delta$ carried by the interaction vertices are absorbed into the canonical normalization.

It remains to include the counterterm contribution. We split it according to the counterterm insertion from which each term arises, see Eq.~(\ref{eq: USR HI}),
\begin{equation}
	\mathcal{P}_{\delta \phi}^{\rm cts} \equiv \mathcal{P}_{\delta \phi}^{\overline{V}_2} + \mathcal{P}_{\delta \phi}^{Z_\phi} \,,
\end{equation}
where using the tadpole-renormalized form of $\overline{V}_2$ in Eq.~(\ref{eq: USR tilde V2}), we find
\begin{align} \label{eq: USR P cts}
	&\mathcal{P}_{\delta\phi}^{\overline V_2}  = \dfrac{k^{3+\delta}}{2 \pi^2} \int \dfrac{\dd^{3+\delta} \vb{p}}{(2\pi)^{3}} \, 2\Im{\delta\phi_k^2 \int_{-\infty_-}^\tau \mkern-10mu \dd \tau' \, a^{2} \, \delta\phi_k^{*2} \left(- \dfrac{V_4}{2} \abs{\delta\phi_{c,p}}^2 - \dfrac{V_3}{2 \phi_0'} a^{2+\delta}  \left( \dfrac{\abs{\delta\phi_{c,p}}^2}{ a^{2+\delta} }\right) ' \right) } \,,\\
	&\mathcal{P}_{\delta \phi}^{Z_\phi} = -Z_\phi \dfrac{k^{3+\delta}}{2 \pi^2}\left[  \abs{\delta\phi_k}^2 + 2 \Im{\delta \phi_k^2 \int_{-\infty_-}^\tau \dd \tau' \mu^\delta a^{4+\delta}V_2 \delta \phi_k^{*2}}\right] \,,
\end{align}
where we have used the Wronskian condition $\Im{\delta \phi_k \delta \phi'^*_k} = (2 \mu^\delta a^{2+\delta})^{-1}$.
The first term in $\mathcal{P}_{\delta\phi}^{\overline V_2}$ cancels exactly the quartic-loop contribution $\mathcal{P}_{\delta\phi}^{V_4}$. This cancellation is model independent and holds for arbitrary $\tau$ and $k$.

Following a strategy analogous to that used in Section~\ref{sec: SIGWs}, we now proceed in two steps. First, we analyze the UV region of the loop integrals, $p \in [L,\infty)$, in order to extract both the divergent and finite UV contributions. Second, we combine this result with the counterterms. As noted in \cite{Kristiano:2025ajj}, in the class of scenarios considered here the cancellation of the total UV divergences is already enforced by tadpole renormalization. The present procedure makes this cancellation explicit and yields a finite renormalized one-loop scalar spectrum, with the UV and IR contributions consistently matched, ready to be evaluated in models with a specified background evolution $H(\tau)$. 

\subsubsection{Ultraviolet contribution and renormalized spectrum} \label{sec: USR UV}

We now analyze the UV part of the loop contribution. Using the notation introduced in Section~\ref{sec: SIGWs}, we decompose any loop contribution into UV and IR pieces as $\mathcal{P} \equiv {}^{\rm (UV)}\mathcal{P} + {}^{\rm (IR)}\mathcal{P}$. The UV contribution associated with the counterterm diagram $\mathcal{P}_{\delta\phi}^{\overline V_2}$, together with the quartic loop diagram $\mathcal{P}_{\delta\phi}^{V_4}$, can be written as
\begin{equation} \label{eq: USR P UV v2+v4}
	^{\rm (UV)}\left( \mathcal{P}_{\delta\phi}^{\overline V_2} + \mathcal{P}_{\delta\phi}^{V_4}\right) = \dfrac{k^{3+\delta}}{2 \pi^2} \int_{p>L} \dfrac{\dd^{3+\delta} \vb{p}}{(2\pi)^{3}} \, 2\Im{\delta\phi_k^2 \int_{-\infty_-}^\tau \mkern-10mu \dd \tau' \, a^{2} \, \delta\phi_k^{*2} \left( - \dfrac{V_3 a^{2+\delta}}{2 \phi_0'} \left( \dfrac{ 1}{2 a^{2+\delta} W_p}\right) ' \right) } \,,
\end{equation}
where we have used the WKB expansion of the canonically normalized modes. The loop integral is then reduced to the large-momentum expansion of $W_p$ and can be evaluated directly.

The one-loop contribution associated with two cubic insertions is treated using Eqs.~(\ref{eq: Dim Reg Ints Tempo 2 Vert I})--(\ref{eq: Dim Reg Ints Tempo 2 Vert II}), which allow us to extract the UV part of $\mathcal{P}_{\delta \phi}^{V_3}$ by trading time integrals for time derivatives. This gives
\begin{align}
	\nonumber
	^{\rm (UV)}\mathcal{P}_{\delta\phi}^{V_3,1} =& \frac{k^{3+\delta}}{2\pi^2} \int_{p>L}\frac{\dd^{3+\delta}\vb{p}}{(2\pi)^3} \abs{\delta\phi_k}^2  \sum_{n,m=0}^\infty \dfrac{i^{n-m}}{(p+q)^{2+n+m}}\\
	& \times  \partial_{\tau}^m\left( a^2V_3 \, e^{i(p+q)\tau}\delta\phi_k \delta\phi_{c,p} \delta\phi_{c,q} \right) \,\partial_{\tau}^n \left(  a^2 V_3\, e^{-i(p+q)\tau}\delta\phi^*_k  \delta\phi_{c,p}^*  \delta\phi_{c,q}^* \right)\,, \\ \nonumber
	^{\rm (UV)}\mathcal{P}_{\delta\phi}^{V_3,2} =& \frac{k^{3+\delta}}{2\pi^2} \int_{p>L} \frac{\dd^{3+\delta}\vb{p}}{(2\pi)^3} \Re  \bigg\{ -2\delta\phi_k^2 \sum_{n = 0}^\infty \dfrac{i^{n-1}}{(p+q)^{1+n}}\\
	&  \times  \int_{-\infty_-}^\tau \mkern-10mu \dd \tau' \, \left(a^2 V_3\,  e^{i(p+q)\tau'}\delta\phi^*_k \delta\phi_{c,p} \delta\phi_{c,q}\right) \,  \partial_{\tau'}^n \left( a^2 V_3\,  e^{-i(p+q)\tau'}\delta\phi_k^* \delta\phi_{c,p}^* \delta\phi_{c,q}^*\right) \bigg\} \,.
\end{align}
Since we are only interested in the UV part of the loop, corrections of order $\order{1/L}$ can be neglected. In the large-momentum limit, the loop modes satisfy $\delta\phi_{c,p} = e^{-i p \tau}/\sqrt{2p}$. It then follows that ${}^{\rm (UV)}\mathcal{P}_{\delta\phi}^{V_3,1}$ is UV convergent and contributes only at order $\order{1/L}$. By contrast, in ${}^{\rm (UV)}\mathcal{P}_{\delta\phi}^{V_3,2}$ the $n=0$ term gives a non-negligible UV contribution. Therefore, the UV contribution associated with the pair of cubic insertions reduces to
\begin{align}
	^{\rm (UV)}\mathcal{P}_{\delta\phi}^{V_3} = \frac{k^{3+\delta}}{2\pi^2} \int_{p>L} \frac{\dd^{3+\delta}\vb{p}}{(2\pi)^3} \Re  \bigg\{  \dfrac{i\, \delta\phi_k^2 }{2pq(p+q)}  \int_{-\infty_-}^\tau \mkern-10mu \dd \tau' \, a^4 V_3^2\,  \delta\phi^{*2}_k  \bigg\} \,.
\end{align}
This term is UV divergent and generates a pole in $\delta$.

It is remarkable that the UV divergence in $\delta$ arising from the combined quartic $\mathcal{P}_{\delta\phi}^{V_4}$ and counterterm $\mathcal{P}_{\delta\phi}^{\overline V_2}$ contribution cancels exactly against the UV divergence generated by the cubic diagram $\mathcal{P}_{\delta\phi}^{V_3}$, as pointed out in \cite{Kristiano:2025ajj}. This cancellation is subtle. Tadpole renormalization fixes $\overline V_1$, and differentiating it gives the two contributions displayed in Eq.~(\ref{eq: USR tilde V2}). One of them cancels the quartic loop diagram exactly, including both its divergent and finite parts, while the second produces a UV divergence that cancels the one from the cubic diagram. This second cancellation originates from the next-to-leading term in the large-momentum expansion of the time derivative of the frequency $W_p$ in Eq.~(\ref{eq: USR P UV v2+v4}). That term depends on the effective scalar mass in Eq.~(\ref{eq: USR meff}), and therefore ultimately on $V_2$, whose time derivative is related to $V_3$. This cancellation is crucial, since it removes the need to introduce additional divergent counterterms in order to renormalize the one-loop spectrum within the approximation considered here; and in particular $Z_\phi$ is finite. A Ward-identity interpretation of this cancellation, relating the UV divergences in the spectrum to the tadpole-induced quadratic counterterm, was recently given in \cite{Fang:2026off}.

A further subtlety must also be stressed. The cancellation just described is not exact if $\epsilon$-suppressed terms are retained: UV-divergent $\epsilon$-suppressed remnants survive. This is expected in the present setup, since the fluctuation action used here is not complete at that order. By construction, we have neglected all interactions suppressed by the slow-roll parameter $\epsilon$, namely those induced by the non-dynamical metric fluctuations. To make this point explicit, we now display the UV contribution obtained by using the truncated Hamiltonian in Eq.~(\ref{eq: USR HI}), while retaining $\epsilon$-suppressed terms elsewhere, for instance the $\dot H$ contribution to the effective mass in Eq.~(\ref{eq: USR meff}). One finds:\footnote{To obtain this result, it is necessary to retain the $\sin^\delta\theta$ factor from the Fourier volume element not only in the cubic diagram, where the integrand depends explicitly on $\theta$ through the momentum $q$, but also in the quartic and counterterm diagrams.}
\begin{align}
	\nonumber
	^{\rm (UV)}\left( \mathcal{P}_{\delta\phi}^{\overline V_2} + \mathcal{P}_{\delta\phi}^{V_4} + \mathcal{P}_{\delta\phi}^{V_3}\right)  =& \dfrac{k^3}{8 \sqrt{2} \pi ^4 M_P} \Im\Bigg\lbrace \delta\phi_k^{2} \int_{-\infty_-}^\tau \mkern-10mu \dd \tau' \,\dfrac{ a^4 V_3}{\sqrt{\epsilon}} \delta \phi_k^{*2} \bigg[ - \dfrac{L^2}{a^2} + \dfrac{V_2}{2} -H^2 \\
	&+ \epsilon \dfrac{H^2}{4} \left( \dfrac{2}{\delta} -1+2\log (2kL) \right)\left(-4+2\epsilon -\eta \right)   \bigg]\Bigg\rbrace\,,
\end{align}
where we have defined the second slow-roll parameter $\eta \equiv \dot \epsilon / (H \epsilon)$.

The second line contains a UV divergence, but it is $\epsilon$-suppressed. Consistently with the action used in this section, and because this is the relevant approximation for the class of models under consideration, these terms must be neglected.

We thus conclude that, within the approximation in which the scalar dynamics is accurately described by the potential and lapse- and shift-induced interactions are neglected, renormalizing the one-loop one-point function is sufficient to render the one-loop scalar power spectrum finite, as found in \cite{Kristiano:2025ajj}. No additional divergent counterterms are required beyond those associated with the renormalization of the potential. To complete the calculation, it only remains to include the IR contribution, computed directly in three spatial dimensions with loop integrals restricted to $p \in [0,L]$.\footnote{Unlike the tensor spectrum analyzed in Section~\ref{sec: SIGWs}, the one-loop scalar power spectrum develops IR divergences when one of the internal loop momenta becomes soft. A fully satisfactory treatment of these IR divergences lies beyond the scope of the present work, which focuses on the UV structure of cosmological loops and on the construction of a systematic, model-independent renormalization procedure.} Neglecting $\epsilon$-suppressed corrections, we obtain
\begin{align}
	\nonumber
	\mathcal{P}_{\delta\phi}^{\rm total} =& \lim_{L \to \infty} \Bigg[ \,^{\rm (IR)}\left( \mathcal{P}_{\delta\phi}^{\overline V_2} + \mathcal{P}_{\delta\phi}^{V_4} + \mathcal{P}_{\delta\phi}^{V_3}\right) - L^2 \dfrac{k^3}{8 \sqrt{2} \pi ^4 M_P} \Im\Bigg\lbrace \delta\phi_k^{2} \int_{-\infty_-}^\tau \mkern-10mu \dd \tau' \,\dfrac{ a^2 V_3}{\sqrt{\epsilon}} \delta \phi_k^{*2}\Bigg\rbrace \Bigg] \\ \nonumber
	&+ \dfrac{k^3}{8 \sqrt{2} \pi ^4 M_P} \Im\Bigg\lbrace \delta\phi_k^{2} \int_{-\infty_-}^\tau \mkern-10mu \dd \tau' \,\dfrac{ a^4 V_3}{\sqrt{\epsilon}} \delta \phi_k^{*2} \bigg[\dfrac{V_2}{2} -H^2 \bigg]\Bigg\rbrace \\
	&-Z_\phi \dfrac{k^{3}}{2 \pi^2}\left[  \abs{\delta\phi_k}^2 + 2 \Im{\delta \phi_k^2 \int_{-\infty_-}^\tau \dd \tau' a^{4}V_2 \delta \phi_k^{*2}}\right] \,.
\end{align}
The first line contains the contributions involving loop-momentum integrals. The $L$-dependent term subtracts the UV divergence of these diagrams. In the present approximation, the only such divergence is quadratic. The second line is the finite remnant left by the UV loop contribution together with the counterterms fixed by tadpole renormalization. Finally, the last line contains the undetermined finite contribution controlled by the renormalized parameter $Z_\phi$, which must be fixed by a renormalization condition or, equivalently, by matching to data. This term makes it explicit that the loop contribution alone cannot be assigned a scheme-independent meaning at a given scale. In particular, any claim based only on the loop correction to the scalar spectrum on modes that are already superhorizon before the transient violation of slow-roll --the regime at the center of the discussion initiated by \cite{Kristiano:2022maq}-- would be scheme dependent and therefore not directly physical.

\section{Discussion} \label{sec: Discussion}

The characterization of loop-level cosmological correlators requires a consistent implementation of renormalization, which is essential to obtain an accurate description of loop effects in cases of theoretical or phenomenological interest. Besides, the adequate treatment of divergences is necessary to quantify the regime of validity of perturbation theory, and therefore of the tree-level predictions as the leading approximation. Renormalization must therefore make explicit that divergences are not observable, but scheme-dependent contributions, indistinguishable from a redefinition of the corresponding finite counterterm coefficients. Once 
divergences are removed, any finite remainder that is degenerate with local counterterm insertions is by construction scheme-dependent and cannot be isolated as an intrinsic loop effect. With these premises, we have studied the renormalization of the one-loop primordial tensor and scalar spectra under minimal assumptions of broad applicability.

Renormalizing expectation values in cosmological backgrounds is technically more involved than in flat-space scattering.
In the in-in formalism, loop momentum integrals are tied to time integrals over modes evolving in a time-dependent background and, in schemes such as dimensional regularization, consistency requires continuing not only the loop measure but also the theory itself away from three spatial dimensions. The progress made in understanding the UV structure of in-in loops in \cite{Ballesteros:2024cef,Ballesteros:2025nhz} allows us to characterize the divergent part of the loops before specifying the background dynamics, using only the universal high-momentum WKB expansion of the internal modes running in the loops. This leads to a direct renormalization strategy: once the divergent structure has been identified, it can be matched to local counterterm insertions and canceled by fixing the divergent parts of the corresponding counterterm coefficients. The result is a finite and renormalized observable whose form remains model independent until a particular background evolution and a concrete field dynamics are specified.

We have applied this strategy first to the tensor spectrum induced at one loop by the fluctuations of a minimally coupled spectator scalar. The calculation was performed on an arbitrary FLRW background and for an arbitrary time-dependent spectator mass, neglecting the interactions mediated by the algebraic metric variables in the spectator regime, with the only additional assumption that the loop modes admit an asymptotic past positive-frequency behavior. The result is a finite and renormalized tensor spectrum that can be applied directly once a concrete background and spectator dynamics are specified. The counterterms required for renormalization descend from the covariant gravitational EFT. They are therefore not an arbitrary local basis for $h_{ij}$. General covariance fixes their structure and, in particular, forbids a tensor-mass counterterm. The fact that the explicit UV divergences of the scalar loop are nevertheless canceled by this restricted covariant counterterm set provides a direct consistency check of the renormalization procedure in a time-dependent background.

As an analytically tractable application, we evaluated the renormalized tensor spectrum in the displaced-mass example in exact de Sitter. In this case, the effective scalar mass corresponds to the massless de Sitter profile shifted by a time scale $\tau_*$. This example illustrates how the general result can be used once a concrete scalar dynamics is chosen, leading to a loop spectrum with distinguishable features. The scale $\tau_*$ generates an intrinsic dependence on the dimensionless combination $\kappa=k\tau_*$, whereas the late-time limit is finite and does not produce a logarithmic growth of the form $\log(-k\tau)$, see \cite{Weinberg:2006ac,Pimentel:2012tw}. The resulting $\kappa$-dependence cannot be absorbed into any combination of finite local counterterms, and therefore represents a scheme-independent loop contribution. Let us stress, however, that the mere presence of an additional scale is not sufficient to guarantee a distinguishable loop effect, as pointed out in \cite{Ballesteros:2025nhz}. If such a scale enters the loop correction multiplied by a perturbatively small coefficient that can also appear in the counterterms, the effect may still be counterterm-degenerate. The displaced-mass example avoids this situation because $\tau_*$ is part of the free scalar dynamics and is not tied to a small interaction coefficient.

We then considered the scalar power spectrum generated by self-interactions coming from its potential. In this part of the analysis, we neglected the interactions mediated by the algebraic metric variables. This is appropriate in the class of inflationary scenarios in which $\epsilon\ll1$ is suppressed with respect to the higher slow-roll parameters --as in inflationary ultra slow-roll-like phases-- where the potential gives the dominant scalar self-interactions \cite{Maldacena:2002vr,Ballesteros:2024zdp}. As in the previously discussed tensor case, the only additional assumption needed for the renormalization procedure is that the loop modes admit an asymptotic past positive-frequency behavior. In this setting, an important ingredient is tadpole renormalization \cite{Kristiano:2025ajj,Inomata:2026csq,Inomata:2025pqa}. Once a background evolution $H(\tau)$ has been chosen, the fluctuation of the scalar field must have vanishing expectation value; otherwise its zero mode would shift the background around which the perturbative expansion is being performed. At tree level this is reflected in the fact that the bulk action for fluctuations starts at quadratic order. At loop level, however, the one-point function is generically generated radiatively, and the condition $\expval{\delta\phi}=0$ fixes the linear potential counterterm required to keep the background $H(\tau)$ fixed.
 
This condition determines the tower of potential counterterms and, in particular, the quadratic potential-like counterterm entering the one-loop scalar spectrum. Within the leading approximation in $\epsilon$, this is sufficient to render the one-loop scalar power spectrum UV finite. No additional divergent counterterms are required beyond those associated with the renormalization of the potential, in agreement with \cite{Kristiano:2025ajj}. 
More specifically, one contribution to the quadratic counterterm fixed by the tadpole cancels the quartic loop diagram exactly, including both its divergent and finite parts, while a second contribution cancels the UV divergence of the cubic diagram. Finite contributions from local counterterms not fixed by this condition, in particular those associated with the scalar field renormalization, remain free and must be specified by renormalization conditions or, equivalently, by matching to observations. This explicit dependence on finite counterterms prevents statements about the size of the scalar loop contribution on CMB scales (such as the original claim of \cite{Kristiano:2022maq}) from being interpreted as scheme-independent statements whenever the loop result is degenerate with local counterterm insertions.

The two cases we have explored provide a concrete implementation of a model-independent route to the renormalization of cosmological one-loop observables. Beyond the particular spectra studied here, the main outcome is a practical strategy for isolating UV divergences and matching them to local counterterms. This strategy could be directly extended to other cosmological correlators. Importantly, the renormalized expressions that we have presented for tensor and scalar spectra can be used for numerical studies of models of phenomenological interest.

\newpage

\appendix

\mysection{Acknowledgments}

{\small
	Work funded by the following grants: PID2021-124704NB-I00 funded by MCIN/AEI/10.13039 /501100011033 \sloppy and by ERDF A way of making Europe, CNS2022-135613 MICIU/AEI/10.13039/501100011033 and by the European Union NextGenerationEU/PRTR, and Centro de Excelencia Severo Ochoa CEX2020-001007-S funded by MCIN/AEI/10.13039/501100011033.
	JGE is supported by a PhD contract {\it contrato predoctoral para formaci\'on de doctores} (PRE2021-100714) associated to the aforementioned Severo Ochoa grant, CEX2020-001007-S-21-3. FR is supported by the research grant number 20227S3M3B “Bubble Dynamics in Cosmological
	Phase Transitions” under the program PRIN 2022 of the Italian Ministero dell’Università e della Ricerca
	(MUR).
}

\section{Counterterm action for the tensor power spectrum} \label{app: Cts action SIGWS}

In this appendix we derive the counterterm action for the scalar-induced tensor spectrum reported in Eq.~(\ref{eq: Cts Action SIGWs}). We work with the metric (and its inverse)
\begin{align}
	g_{\mu \nu}=\left(\begin{array}{cc}
		-1 & 0 \\
		0 & a^2 \tg_{i j}
	\end{array}\right)  \quad {\rm and} \quad 
	g^{\mu \nu}=\left(\begin{array}{cc}
		-1 & 0 \\
		0 & a^{-2} \tg^{i j}
	\end{array}\right) \,,
\end{align}
which corresponds to the ADM line element Eq.~(\ref{eq: ADM ds2}) with unit lapse and vanishing shift, $N=1$ and $N_i=0$. In the gauge of interest, only tensor fluctuations are relevant and $\tg_{ij} = (e^h)_{ij}$, with $h_{ij}$ transverse and traceless. We keep $d$ spatial dimensions throughout, as required for a consistent derivation of counterterms within dimensional regularization.

Our goal is to determine, up to quadratic order in $h_{ij}$, the curvature invariants $R$, $R_{\mu\nu}R^{\mu\nu}$ and $R_{\mu\nu\rho\sigma}R^{\mu\nu\rho\sigma}$ entering Eq.~(\ref{eq:Cts action general}). We start from the Christoffel symbols, $\Gamma^\mu_{\nu\rho} = \frac{1}{2}g^{\mu\sigma}\left(2 \partial_{(\nu} g_{\rho) \sigma} - \partial_\sigma g_{\nu \rho} \right) $, whose non-vanishing components are
\begin{equation}
	\Gamma^0_{ij} = \dfrac{a^2}{2}\left(2H \tg_{ij} + \dot{\tg}_{ij} \right) \,, \quad \Gamma^i_{0j} = H \tensor{\delta}{^i_j} +  \dfrac{1}{2} \tg^{il} \dot \tg_{lj} \quad {\rm and} \quad \Gamma^i_{jk} = \,^{(d)}\Gamma^i_{jk} \,,
\end{equation}
where $\,^{(d)}\Gamma^i_{jk}$ is the Christoffel symbol associated with the spatial metric $\tg_{ij}$, and dots denote derivatives with respect to cosmic time $t$.

Using $\tensor{R}{^\rho_\sigma_\mu_\nu} = 2 \partial_{[\mu} \Gamma^\rho_{\nu]\sigma} + 2 \Gamma^\rho_{\lambda [\mu} \Gamma^\lambda_{\nu] \sigma}$, it is convenient to work with $R_{\rho\sigma\mu\nu} = g_{\rho \alpha} \tensor{R}{^\alpha_\sigma_\mu_\nu}$, which satisfies
\begin{equation}
	R_{\rho\sigma\mu\nu} = - R_{\sigma\rho\mu\nu} = - R_{\rho\sigma\nu\mu} = R_{\mu\nu\rho\sigma} \,.
\end{equation}
The independent non-vanishing components of the Riemann tensor are then
\begin{align}
	R_{0i0j} &= -\dfrac{a^2}{2}\left(\ddot{\tg}_{ij} + 2H \dot \tg_{ij} - \dfrac{1}{2} \tg^{kl} \dot\tg_{il} \dot\tg_{jk} + 2(\dot H + H^2) \tg_{ij} \right) \,, \\
	R_{0ijk} &= a^2\left( \partial_{[k} \dot \tg_{j]i} + \dot\tg_{l[k} \,^{(d)} \Gamma^l_{j]i} \right) \,, \quad R_{ijkl} = a^2 \,^{(d)}R_{ijkl} + \dfrac{a^4}{2}\left(2H\tg_{i[k} + \dot\tg_{i[k} \right) \left(2H\tg_{l]j} + \dot\tg_{l]j} \right) \,,
\end{align}
where $\,^{(d)}R_{ijkl}$ is the Riemann tensor built from $\tg_{ij}$.

Contracting indices,  $R_{\mu\nu} = g^{\alpha\beta}R_{\alpha\mu\beta\nu} = R_{\nu\mu}$, we obtain
\begin{align} 
	R_{00} &= -\dfrac{1}{2} \tg^{kl}\left(\ddot \tg_{kl} + 2H \dot\tg_{kl} \right) - \dfrac{1}{4} \dot\tg^{kl} \dot\tg_{kl} - d(\dot H + H^2) \,, \quad R_{0i} = \tg^{kl} \left( \partial_{[l} \dot \tg_{i]k} + \dot\tg_{n[l} \,^{(d)} \Gamma^n_{i]k} \right) \,,\\
	R_{ij} &= \,^{(d)} R_{ij} + \dfrac{a^2}{2}\left(\ddot\tg_{ij} + d\, H \dot \tg_{ij} - \tg^{kl} \dot\tg_{il} \dot\tg_{jk} + 2(\dot H +d \, H^2) \tg_{ij} + H \tg_{ij} \tg^{kl}\dot\tg_{kl}  + \dfrac{1}{2} \dot\tg_{ij} \tg^{kl} \dot \tg_{kl} \right) \,. 
\end{align}
Finally, the Ricci scalar $R=g^{\mu\nu}R_{\mu\nu}$ is
\begin{equation}
	R = a^{-2} \,^{(d)} R + \tg^{\ij} \ddot \tg_{ij} + (d+1)H \tg^{ij} \dot \tg_{ij} + \dfrac{3}{4} \dot\tg^{ij} \dot\tg_{ij} + \dfrac{1}{4}\left(\tg^{ij} \dot\tg_{ij} \right) ^2 + d(2\dot H + (d+1)H^2) \,.
\end{equation}

These expressions simplify substantially because $\tg^{ij} \dot \tg_{ij} = \partial_t \log \tg = 0$ since $\tg_{ij}=(e^{h})_{ij}$ with $h_{ij}$ traceless implies $\tg = \det \tg_{ij} = e^{\tr h}=1$. Therefore,
\begin{align} 
	R_{00} &= \dfrac{1}{4} \dot\tg^{kl} \dot\tg_{kl} - d(\dot H + H^2) \,, \quad R_{0i} = \tg^{kl} \left( \partial_{[l} \dot \tg_{i]k} + \dot\tg_{n[l} \,^{(d)} \Gamma^n_{i]k} \right) \,,\\
	R_{ij} &= \,^{(d)} R_{ij} + \dfrac{a^2}{2}\left(\ddot\tg_{ij} + d\, H \dot \tg_{ij} - \tg^{kl} \dot\tg_{il} \dot\tg_{jk} + 2(\dot H +d \, H^2) \tg_{ij}  \right) \,. 
\end{align}
and
\begin{equation}
	R = a^{-2} \,^{(d)} R - \dfrac{1}{4} \dot\tg^{ij} \dot\tg_{ij} + d(2\dot H + (d+1)H^2) \,.
\end{equation}

To renormalize the tensor spectrum at one-loop we only need the pieces of the curvature invariants that are quadratic in $h_{ij}$.\footnote{A scalar cannot be built from a single insertion of $h_{ij}$ being transverse and traceless. Therefore the curvature invariants start at quadratic order in $h_{ij}$.}
Keeping only the quadratic terms, we find
\begin{align}
	R \supset& \dfrac{1}{4}\left( \dot h^2_{ij} - \dfrac{1}{a^2} (\partial_k h_{ij})^2\right)  \,,\\
	R^2 \supset& \dfrac{d(2\dot H + (d+1)H^2)}{2}\left( \dot h^2_{ij} - \dfrac{1}{a^2} (\partial_k h_{ij})^2\right) \,,\\
	R_{\mu\nu}^2 \supset& \dfrac{1}{4} (\ddot h_{ij} + d\, H \dot h_{ij} - a^{-2} \partial^2h_{ij})^2 + \dfrac{d(\dot H + H^2)}{2} \dot h_{ij}^2 -\dfrac{\dot H + d\, H^2}{2a^2} (\partial_k h_{ij})^2 \,,\\ \nonumber
	R_{\mu\nu\rho\sigma}^2 \supset& (\ddot h_{ij} + 2 H \dot h_{ij})^2 + (2 \dot H + (d-1) H^2) \dot h_{ij}^2 - \dfrac{2}{a^2}(\partial_k \dot h_{ij})^2 - \dfrac{H^2}{a^2} (\partial_k h_{ij})^2 + \dfrac{1}{a^4}(\partial^2h_{ij})^2 - \dfrac{2 H}{a^2} \dot h_{ij} \partial^2h_{ij} \\
	& + \dfrac{2}{a^2} \partial_j \dot h_{ik} \partial_k \dot h_{ij} + \dfrac{1}{a^4} \partial_{ij} h_{kl} \partial_{kl} h_{ij} - \dfrac{2}{a^4} \partial_{jl} h_{ik} \partial_{kl} h_{ij}\,.
\end{align}
Although these expressions look cumbersome, their contribution to the counterterm action can be simplified using two facts: (i) purely spatial total derivatives do not affect correlators in the in-in formalism and can be dropped, and (ii) counterterms enter the in-in computation as interaction vertices. Therefore, in the interaction picture the tensor fields appearing in the counterterms satisfy the free equation of motion,
\begin{equation} \label{eq: App cts eom h}
	\ddot h_{ij} + d\, H \dot h_{ij} - a^{-2} \partial^2h_{ij} = 0\,.
\end{equation}

A further subtlety is that, in general, the interaction Hamiltonian in the interaction picture $H_I$ does not coincide with $-L_I$ (see e.g. \cite{Pimentel:2012tw,Ballesteros:2025nhz}), but the corrections to $H_I=-L_I$ are suppressed by the normalization of the free action and hence by $M_P^{-2}$. As explained in Section~\ref{sec: SIGWs}, the counterterm diagrams relevant for the one-loop renormalization of the tensor spectrum cannot carry any additional $M_P$ suppression beyond that already present in the free fields. We may therefore set $H_I=-L_I$ for the present purpose.

Finally, the mixed term proportional to $H \dot h_{ij}\partial^2 h_{ij}$ in $R_{\mu\nu\rho\sigma}^2$ can be traded for a temporal boundary term plus bulk terms by integration by parts:
\begin{equation} \label{eq: App cts bt time}
	\int \dd t \dd^d \vb{x} \, 2 H a^{d-2} \dot h_{ij} \partial^2 h_{ij} = \int \dd t \dd^d \vb{x} \, \partial_t (H a^{d-2} h_{ij} \partial^2 h_{ij}) + a^{d-2} (\dot H + (d-2)H^2) (\partial_k h_{ij})^2 + \partial_i (\mydots) \,.
\end{equation}
Here we used $\sqrt{-g}=a^d$, which is independent of $h_{ij}$ because the tensor is traceless, and we dropped total spatial derivatives. While total time derivatives do not vanish identically in the in-in formalism, the boundary term in Eq.~(\ref{eq: App cts bt time}) does not contribute to the tensor two-point function because it depends only on $h_{ij}$ and not on its time derivatives \cite{Arroja:2011yj,Burrage:2011hd,Braglia:2024zsl}. Equivalently, for a single insertion of the interaction Hamiltonian as in Eq.~(\ref{eq: In-In short 1}), the time integral collapses to the upper limit (the lower limit vanishes due to the $i\epsilon$ prescription), and the result becomes the imaginary part of a real expectation value, which vanishes. We therefore drop this temporal boundary term in practice.

Implementing these simplifications (dropping spatial boundary terms, using Eq.~(\ref{eq: App cts eom h}), and applying Eq.~(\ref{eq: App cts bt time})), the relevant quadratic pieces reduce to
\begin{align}
	R^2 \supset& \dfrac{d(2\dot H + (d+1)H^2)}{2}\left( \dot h^2_{ij} - \dfrac{1}{a^2} (\partial_k h_{ij})^2\right) \,, \\
	R_{\mu\nu}^2 \supset& \dfrac{d(\dot H + H^2)}{2} \dot h_{ij}^2 -\dfrac{\dot H + d\, H^2}{2a^2} (\partial_k h_{ij})^2 \,,\\ \nonumber
	R_{\mu\nu\rho\sigma}^2 \supset& \dfrac{2}{a^4}\left( (\partial^2h_{ij})^2 - a^2 (\partial_k\dot h_{ij})^2\right) + 2 \dot H\left( \dot h_{ij}^2 -\dfrac{d-1}{2a^2} (\partial_k h_{ij})^2\right) \\
	&+ H^2(d^2-3d+3)\left(\dot h_{ij}^2 - a^{-2}(\partial_kh_{ij})^2 \right) \,.
\end{align}

A useful observation is that, if $\dot H=0$, the structure generated by $R^2$ can be used to absorb the contributions from $R_{\mu\nu}^2$ as well as the $H^2$-dependent piece of $R_{\mu\nu\rho\sigma}^2$. This motivates choosing linear combinations that make the independent operator basis as simple as possible. Restricting to these three curvature invariants, the most general independent action can be written as
\begin{align} \label{eq: App cts Scts I}
	\nonumber
	S = \int \dd t \dd^d \vb{x} \, a^d \, \Bigg(& C_1 \left( \dfrac{2}{d+1}\dot H + H^2\right) \left( \dot h^2_{ij} - \dfrac{1}{a^2} (\partial_k h_{ij})^2\right) + C_2 \dot H \left( \dot h^2_{ij} + \dfrac{1}{d\, a^2} (\partial_k h_{ij})^2\right) \\ 
	&+C_3\left[\dfrac{1}{a^4}\left( (\partial^2h_{ij})^2 - a^2 (\partial_k\dot h_{ij})^2\right) -\dfrac{(d-1)(d-4)}{2} \dot H \dot h_{ij}^2 \right] \Bigg) \,.
\end{align}
The coefficient $C_1$ is associated with the $R^2$ operator, up to an overall normalization. The coefficient $C_2$ multiplies the specific linear combination
\begin{equation}
	\frac{2 ((d+1) R_{\mu\nu}^2-R^2)}{(d-1) d} \,,
\end{equation}
whereas $C_3$ corresponds to the structure
\begin{equation}
	\frac{(6-4 d) R^2-2 d (d(d-6) +7) R_{\mu\nu}^2+d(d-1)  R_{\mu\nu\rho\sigma}^2}{2 d (d-1) } \,.
\end{equation}

The counterterm action in Eq.~(\ref{eq:Cts action general}) also includes the non-minimal coupling to the scalar. Expanding it to quadratic order in tensor fluctuations gives
\begin{equation}
	\tilde V''(\overline{\chi})R \supset  \dfrac{\tilde{m}_\chi^2(t)}{4}\left( \dot h^2_{ij} - \dfrac{1}{a^2} (\partial_k h_{ij})^2\right) \,, \quad {\rm with} \quad \tilde{m}_\chi^2(t) \equiv \tilde V''(\chi_0) \,.
\end{equation}
Finally, for the $\Box R$ operator we use
\begin{equation}
	\sqrt{-g} \, \Box R = \partial_\mu(\sqrt{-g}g^{\mu\nu}\partial_\nu R) \supset -\dfrac{1}{4} \partial_t\left( a^d\partial_t \left( \dot h^2_{ij} - \dfrac{1}{a^2} (\partial_k h_{ij})^2\right)\right) +\partial_i(\mydots) \,.
\end{equation}
Repeating the same steps as before --taking time derivatives, using Eq.~(\ref{eq: App cts eom h}), dropping spatial boundary terms, and applying Eq.~(\ref{eq: App cts bt time})-- one finds that many pieces are not independent and can be re-expressed in terms of the operators already present in Eq.~(\ref{eq: App cts Scts I}). Including the independent contribution leads to the final counterterm action
\begin{align} \label{eq: App cts Scts II}
	\nonumber
	&S = \int \dd t \dd^d \vb{x} \, a^d \, \Bigg\lbrace \left[ C_1 \left( \dfrac{2}{d+1}\dot H + H^2\right) + \dfrac{\tilde m_\chi^2(t)}{4}\right]  \left( \dot h^2_{ij} - \dfrac{1}{a^2} (\partial_k h_{ij})^2\right) + C_2 \dot H \left( \dot h^2_{ij} + \dfrac{1}{d\, a^2} (\partial_k h_{ij})^2\right) \\ 
	&+C_3\left[\dfrac{1}{a^4}\left( (\partial^2h_{ij})^2 - a^2 (\partial_k\dot h_{ij})^2\right) -\dfrac{(d-1)(d-4)}{2} \dot H \dot h_{ij}^2 \right] +C_4\left(\dfrac{d-1}{d}\dot H + H^2 \right) \dfrac{1}{a^2}(\partial_kh_{ij})^2 \Bigg\rbrace \,.
\end{align}
The role of the $\Box R$ term, encoded in $C_4$, is to break the apparent degeneracy of the combination $\dot h_{ij}^2 - a^{-2}(\partial_k h_{ij})^2$, even when $\dot H = 0$. We also recall that the coefficients $C_i$ are dimensionless, whereas $\tilde m_\chi^2(t)$ has the interpretation of a time-dependent mass-squared associated to the non-minimal coupling.

Although ultimately we are interested in three spatial dimensions, $d=3$, the dimensional-regularization framework requires tracking the dependence on $d=3+\delta$. Even though counterterms do not contain loop integrals themselves, their coefficients absorb the UV divergences of the loops and are therefore formally divergent, containing poles in $\delta$:
\begin{equation}
	C_i \equiv \dfrac{1}{\delta} C_{d,i} + C_{f,i}  \quad {\rm and} \quad \tilde{m}_\chi^2(t) \equiv \dfrac{1}{\delta} \tilde{m}_{d,\chi}^2(t) + \tilde{m}_{f,\chi}^2(t) \,.
\end{equation}
As a consequence, $\order{\delta}$ corrections from the operators can combine with the $1/\delta$ poles in the couplings to produce finite contributions. Expanding Eq.~(\ref{eq: App cts Scts II}) around $d=3+\delta$ and keeping the terms that can generate such finite remnants yields
\begin{align} \label{eq: App cts Scts III}
	\nonumber
	S = &\int \dd t \dd^d \vb{x} \, a^d \, \Bigg\lbrace \left[ C_1 \left( \dfrac{1}{2}\dot H + H^2\right) + \dfrac{\tilde m_\chi^2(t)}{4}\right]  \left( \dot h^2_{ij} - \dfrac{1}{a^2} (\partial_k h_{ij})^2\right) + C_2 \dot H \left( \dot h^2_{ij} + \dfrac{1}{3\, a^2} (\partial_k h_{ij})^2\right) \\ \nonumber
	&+C_3\left[\dfrac{1}{a^4}\left( (\partial^2h_{ij})^2 - a^2 (\partial_k\dot h_{ij})^2\right) + \dot H \dot h_{ij}^2 \right] +C_4\left(\dfrac{2}{3}\dot H + H^2 \right) \dfrac{1}{a^2}(\partial_kh_{ij})^2  \\
	&+\delta \, \dot H \,  \Bigg[  - \dfrac{C_1}{8} \left( \dot h^2_{ij} - \dfrac{1}{a^2} (\partial_k h_{ij})^2\right) -\dfrac{C_3}{2} \dot h_{ij}^2  +\dfrac{C_4 - C_2}{9\, a^2}(\partial_kh_{ij})^2 \Bigg] \Bigg\rbrace+ \order{\delta} \,.
\end{align}
Crucially, the explicit extra-dimensional corrections from the curvature invariants (the last line above) are weighted by $\dot H$. Therefore, they only play a role beyond strictly constant-$H$ backgrounds. We neglect the $\order{\delta^2}$ corrections  since combined with the $1/\delta$ poles in the couplings they would only produce $\order{\delta}$ contributions, which vanish in the limit $\delta\to 0$.

\section{Intermediate UV results for the tensor power spectrum} \label{app: formulas}

In this appendix we collect the intermediate UV expressions entering the one-loop scalar-induced tensor spectrum. These formulas are used in Section~\ref{sec: SIGWs} to obtain the divergent and finite UV contributions before adding the counterterms. All background quantities and mode functions inside time integrals are evaluated at the integration time $\tau'$.

The UV contribution of the quartic diagram in Eq.~(\ref{eq: Ph q UV}) takes the form
\begin{equation} \label{eq: Ph q UV sol}
	\prescript{\rm (UV)}{}{\mathcal{P}}_h^{\rm q} = \frac{1}{\delta} \prescript{\rm (UV)}{d}{\mathcal{P}}_h^{\rm q} + \prescript{\rm (UV)}{f}{\mathcal{P}}_h^{\rm q}\,,
\end{equation}
with
\begin{align}
	\nonumber
	&\prescript{\rm (UV)}{d}{\mathcal{P}}_h^{\rm q} = \frac{k^3}{192 \pi^4} \Im\Bigg\lbrace h_k^2 \int_{-\infty_-}^\tau \dd \tau' \,  a^2 h_k^{*2}\Big( -4 a H (m^2_\chi)'-(m^2_\chi)'' \\
	&\quad+ a^2 \left(\dddot{H}+9 \ddot{H} H+2 m^2_\chi \left(2 \dot{H}+3 H^2\right)+3 \dot{H} \left(\dot{H}+6 H^2\right)-3 m^4_\chi\right)\Big) \Bigg\rbrace \,,
\end{align}
and
\begin{align}
	\nonumber
	&\prescript{\rm (UV)}{f}{\mathcal{P}}_h^{\rm q} = \prescript{\rm (UV)}{d}{\mathcal{P}}_h^{\rm q} \, \log (2kL) + \frac{k^3}{192 \pi^4} \Im \Bigg\lbrace h_k^2 \int_{-\infty_-}^\tau \dd \tau' \, a^2 h_k^{*2}\Big(\frac{1}{12} \Big(a^2 \Big(5 \dddot{H}+57 \ddot{H} H+2 m^2_\chi \left(16 \dot{H}+51 H^2\right)\\
	&+9 \left(6 \dot{H} H^2+\dot{H}^2-12 H^4\right)+3 m^4_\chi\Big)+4 a H (m^2_\chi)'+(m^2_\chi)''\Big)-\frac{2 L^4}{a^2}-2 L^2 \left(\dot{H}+2 H^2-m^2_\chi\right) \Big) \Bigg\rbrace \,.
\end{align}

The first cubic contribution, defined in Eq.~(\ref{eq: Ph c1 UV}), gives
\begin{equation} \label{eq: Ph c1 UV sol}
	\prescript{\rm (UV)}{}{\mathcal{P}}_h^{\rm c,1} = \frac{1}{\delta} \prescript{\rm (UV)}{d}{\mathcal{P}}_h^{\rm c,1} + \prescript{\rm (UV)}{f}{\mathcal{P}}_h^{\rm c,1}\,,
\end{equation}
where
\begin{align}
	\prescript{\rm (UV)}{d}{\mathcal{P}}_h^{\rm c,1} = - \frac{k^3 \abs{h_k}^2}{1920 \pi^4}  \Im{ 2 h_k h'^*_k \left(k^2-5 a^2 \left(\dot{H}+2 H^2-m^2_\chi\right)\right)+ h'^*_k h''_k +h_k h'''^*_k } \,,
\end{align}
and
\begin{align}
	\nonumber
	&\prescript{\rm (UV)}{f}{\mathcal{P}}_h^{\rm c,1} = \prescript{\rm (UV)}{d}{\mathcal{P}}_h^{\rm c,1} \, \log (2kL) + \frac{k^3 \abs{h_k}^2}{3840 \pi^4} \Bigg(-\frac{8}{3} L^3 \abs{h_k}^2 +4  L^2 \Im{ h_k h'^*_k} \\ \nonumber
	&\quad+L \Big(  \frac{2}{7} \abs{h_k}^2 \left(13 k^2-56 a^2 \left(\dot{H}+2 H^2-m^2_\chi\right)\right)+4 \Re{ h_k h''^*_k}-2 \abs{h'_k}^2\Big)  \\
	&\quad+\frac{1}{210} \Im{ 2 h_kh'^*_k \left(35 a^2 \left(13 \dot{H}+56 H^2+17 m^2_\chi\right)+29 k^2\right)+119 h'^*_k h''_k +119 h_k h'''^*_k} \Bigg) \,.
\end{align}

The second cubic contribution, defined in Eq.~(\ref{eq: Ph c2 UV}), gives
\begin{equation} \label{eq: Ph c2 UV sol}
	\prescript{\rm (UV)}{}{\mathcal{P}}_h^{\rm c,2} = \frac{1}{\delta} \prescript{\rm (UV)}{d}{\mathcal{P}}_h^{\rm c,2} + \prescript{\rm (UV)}{f}{\mathcal{P}}_h^{\rm c,2}\,,
\end{equation}
with
\begin{align}
	\nonumber
	&\prescript{\rm (UV)}{d}{\mathcal{P}}_h^{\rm c,2} = \frac{k^3}{1920 \pi^4} \Im\Bigg\lbrace h_k^2 \int_{-\infty_-}^\tau \dd \tau'  \Big(h^*_kh''''^*_k+2 k^2h^*_k h''^*_k \\ \nonumber
	&\quad + 10 a^2 h^*_k \Big(h'^*_k \left((m^2_\chi)'-a \left(\ddot{H}+6 \dot{H} H-2 H m^2_\chi+4 H^3\right)\right)+h''^*_k \left(-\dot{H}-2 H^2+m^2_\chi\right)\Big) \\ \nonumber
	&\quad+h^{*2}_k \Big(k^4-10 a^2 \Big(a^2 \left(\dddot{H}+9 \ddot{H} H+2 m^2_\chi \left(2 \dot{H}+3 H^2\right)+3 \dot{H} \left(\dot{H}+6 H^2\right)-3 m^4_\chi\right)\\
	&\qquad \qquad-4 a H (m^2_\chi)'+k^2 \left(\dot{H}+2 H^2-m^2_\chi\right)-(m^2_\chi)''\Big)\Big)\Big)  \Bigg\rbrace \,,
\end{align}
and
\begin{align}
	\nonumber
	&\prescript{\rm (UV)}{f}{\mathcal{P}}_h^{\rm c,2} = \prescript{\rm (UV)}{d}{\mathcal{P}}_h^{\rm c,2} \, \log (2kL) -\frac{k^3}{1920 \pi^4} \Im\Bigg\lbrace h_k^2 \int_{-\infty_-}^\tau \dd \tau' h^*_k \Big( -4 L^4 h^{*}_k - i \frac{8}{3} L^3 h'^*_k\\ \nonumber
	&\quad +\frac{2}{7} L^2 \left(h^*_k \left(11 k^2-42 a^2 \left(\dot{H}+2 H^2-m^2_\chi\right)\right)+7 h''^*_k\right)\\ \nonumber
	&\quad+ i \frac{2}{7} L \left(-28 a^2 \left(a h^*_k \left(\ddot{H}+6 \dot{H} H-2 H m^2_\chi+4 H^3 - \frac{(m^2_\chi)'}{a}\right)+2 h'^*_k \left(\dot{H}+2 H^2-m^2_\chi\right)\right)+7 h'''^*_k+13 k^2 h'^*_k\right)\\ \nonumber
	&\quad+ \frac{1}{420} \Big(h^*_k \Big(10 a^2 \Big(7 a^2 \left(13 \dddot{H}+177 \ddot{H} H+2 m^2_\chi \left(56 \dot{H}+219 H^2\right)+9 \left(6 \dot{H} H^2+\dot{H}^2-60 H^4\right)+51 m^4_\chi\right) \\ \nonumber
	&\qquad \qquad+476 a H (m^2_\chi)'+k^2 \left(211 \dot{H}+632 H^2-m^2_\chi\right)+119 (m^2_\chi)''\Big)-141 k^4\Big)\\ \nonumber
	&\qquad \qquad+70 a^2 \left(h'^*_k \left(a \left(13 \ddot{H}+2 H \left(69 \dot{H}+56 H^2+17 m^2_\chi\right)\right)+17 (m^2_\chi)'\right)+h''^*_k \left(13 \dot{H}+56 H^2+17 m^2_\chi\right)\right)\\
	&\qquad \qquad+119 h''''^*_k+58 k^2 h''^*_k\Big) \Big) \Bigg\rbrace \,.
\end{align}

Adding the quartic and cubic diagrams, and simplifying the result with the tensor equation of motion in $3+\delta$ spatial dimensions together with repeated integrations by parts, the total divergent UV contribution can be written as
\begin{equation}
	\prescript{\rm (UV)}{d}{\mathcal{P}}_h^{\rm loop}(\tau,k) = \frac{k^3}{960 \pi ^4 } \Im \Bigg\{ h_k^2 \int_{-\infty_-}^\tau \dd \tau'  a^2 \, \bigg[ 2 \dot H  k^2 h_k^{*2} + (h'^{*2}_k-k^2 h_k^{*2}) \left(6 \dot H+9 H^2-5 m_\chi^2 \right)\bigg]\Bigg\} \,.
\end{equation}
The corresponding finite UV contribution is
\begin{align} \label{eq: Ph loop UV f sol}
	\nonumber
	&\prescript{\rm (UV)}{f}{\mathcal{P}}_h^{\rm loop}(\tau,k) = \log 2kL \, \prescript{\rm (UV)}{d}{\mathcal{P}}_h^{\rm loop}(\tau,k)  -L \frac{k^3 \Im{h_kh'^*_k}^2}{960 \pi ^4} \\ \nonumber
	& \qquad - L^2\dfrac{k^3}{120 \pi^4} \Im \Bigg\{ h_k^2 \int_{-\infty_-}^\tau \dd \tau' \bigg[ L^2 h_k^{*2}  +\frac{1}{56}  \left(h_k^{*2} \left(28 a^2 \left(\dot H +2 H^2-m^2_\chi \right)+11 k^2\right)-7 h_k'^{*2}\right) \bigg]  \Bigg\}\\ \nonumber
	& \qquad +\frac{k^3}{403200 \pi ^4} \Im\Bigg\{  h_k^2 \int_{-\infty_-}^\tau \dd \tau' \, \bigg[  h_k ^{*2} \Big[a^2 \left(420 a^2 \left(\dddot H +9 \ddot H   H +2 m_\chi^2 \left(2 \dot H +3  H ^2\right) \right.\right. \\ \nonumber
	& \qquad \left.\left.+3 \dot H  \left(\dot H +6  H ^2\right)-3 m_\chi^4\right)-1680 a  H  (m_\chi^2)'+k^2 \left(-964 \dot H -3069  H ^2+5 m_\chi^2\right)-420 (m_\chi^2)''\right)+130 k^4\Big]\\
	&\qquad+h'^{*2}_k \Big[ 7 a^2 \left(78 \dot H +207  H ^2+85 m_\chi^2\right)-90 k^2\Big]\bigg]\Bigg\} \,.
\end{align}

\subsection{Logarithmic time integrals} \label{app: Log time int}

In the displaced-mass application to the tensor power spectrum, the only time integrals in Eq.~(\ref{eq: SIWGs Final Spectrum}) that are not elementary are those containing the factor $\log(L/(a\mu))$. They can all be written in the form, for $\alpha=0,1,2$,
\begin{align}
	\nonumber
	\int^\tau \dd \tau' \,  e^{2i k \tau'} \, \frac{\tau'^\alpha}{(\tau' - \tau_*)^2} \log \frac{L}{a(\tau') \mu} &= \partial_\eta^\alpha \left[ \int^\tau \dd \tau' \, \frac{e^{2i k \tau' + \eta \tau'}}{(\tau' - \tau_*)^2} \log \frac{L}{a(\tau') \mu}\right] \eval_{\eta = 0} \\
	&= \partial_\eta^\alpha \mathcal{J}(2i k + \eta,\tau,\tau_*) \eval_{\eta = 0} \,,
\end{align}
where
\begin{align}
	\nonumber
	\mathcal{J}(s,\tau,\tau_*) \equiv& - \dfrac{e^{s \tau}}{\tau - \tau_*} \log \frac{L}{a \mu} + s e^{s \tau_*} \left(\log \frac{L}{a \mu} \left( {\rm Ei}(s(\tau-\tau_*))+ i \pi\right)   - \mathcal{E}_s(\tau,\tau_*) \right) \\
	&+ \frac{e^{s\tau_*} {\rm Ei}(s(\tau-\tau_*)) -  {\rm Ei}(s\tau)}{\tau_*} \,,
\end{align}
and
\begin{equation}
	\mathcal{E}_s(\tau,\tau_*) \equiv \int_{-\infty_-}^\tau \dd \tau' \, \frac{{\rm Ei}(s(\tau'-\tau_*)) + i \pi }{\tau'} = \int_{\infty_-}^{-\tau/\tau_*} \dd x' \, \frac{{\rm Ei}(-s\tau_*(1+x')) + i \pi }{x'}  \,,
\end{equation}
where we used $x' = -\tau' / \tau_*$. Since the relevant quantity is $\mathcal{E}_{2ik}(\tau,\tau_*)$, we also introduce $x=-\tau/\tau_*$ and work with
\begin{equation}
	\mathcal{M}(x) \equiv \int_{\infty_-}^{x} \dd x' \dfrac{g(x')}{x'} \, \quad {\rm where} \quad g(x') = {\rm Ei}( - 2i \kappa (1+x')) + i \pi \,.
\end{equation}
Since $g(x')$ is regular at $x'=0$, the integral has a logarithmic late-time behavior,
\begin{align}
	\nonumber
	\mathcal{M}(x) &= \int_{\infty_-}^{1} \dd x' \dfrac{g(x')}{x'} + \int_{1}^{x} \dd x' \dfrac{g(x')-g(0)}{x'} + g(0) \log x \\
	&= \int_{\infty_-}^{1} \dd x' \dfrac{g(x')}{x'} + \int_{1}^{0} \dd x' \dfrac{g(x')-g(0)}{x'} + g(0) \log x + \order{x}\,.
\end{align}
The finite part can be extracted by introducing, for $\alpha>0$,
\begin{align}
	\nonumber
	F_\alpha &\equiv \int_{\infty_-}^{0} \dd x' x'^{\alpha - 1} g(x') = \int_{\infty_-}^{1} \dd x' x'^{\alpha-1} g(x') + \int_{1}^{0} \dd x' x'^{\alpha-1}(g(x')-g(0)) - \frac{g(0)}{\alpha} \\
	&= \int_{\infty_-}^{1} \dd x' \dfrac{g(x')}{x'} + \int_{1}^{0} \dd x' \dfrac{g(x')-g(0)}{x'} -\frac{g(0)}{\alpha} + \order{\alpha} \,.
\end{align}
Using $F_\alpha = (2i \kappa)^{-\alpha} {\rm E}_{1+\alpha}(2i \kappa) \Gamma(\alpha)$ (where $E_n(z)$ denotes the generalized exponential integral), and recalling that $\mathcal{M}(x)$ is defined through $\mathcal{M}(-\tau/\tau_*)=\mathcal{E}_{2ik}(\tau,\tau_*)$, the finite part of $\mathcal{M}(x)$ is obtained from the finite part of $F_\alpha$ in the limit $\alpha\to0$.

\providecommand{\href}[2]{#2}\begingroup\raggedright\endgroup

\end{document}